\documentclass[STIX1COL]{WileyNJD-v2}

\usepackage{moreverb}
\usepackage{graphicx}
\usepackage{listings}
\usepackage{amsmath}
\usepackage{textcomp}
\usepackage[utf8]{inputenc}
\usepackage{tabularx}
\usepackage{xspace}
\usepackage{pifont}
\usepackage{epstopdf}
\usepackage{amssymb}
\usepackage{gensymb}
\usepackage{url}
\usepackage{multirow}
\graphicspath{{./plots/}}
\usepackage{adjustbox,lipsum}
\usepackage{algpseudocode}
\usepackage{listings}
\usepackage[numbers,sort&compress]{natbib}

\newcommand\BibTeX{{\rmfamily B\kern-.05em \textsc{i\kern-.025em b}\kern-.08em
T\kern-.1667em\lower.7ex\hbox{E}\kern-.125emX}}

\articletype{Article Type}%

\received{<day> <Month>, <year>}
\revised{<day> <Month>, <year>}
\accepted{<day> <Month>, <year>}

\begin{document}
\title{Heuristics and Metaheuristics for Dynamic Management of Computing and Cooling Energy in Cloud Data Centers}

\author[1,2]{Patricia Arroba}

\author[2,3]{José L. Risco-Martín}

\author[1,2]{José M. Moya}

\author[2,3]{José L. Ayala}

\authormark{Patricia Arroba \textsc{et al}}

\address[1]{\orgdiv{Laboratorio de Sistemas Integrados (LSI)}, \orgname{Universidad Politécnica de Madrid}, \orgaddress{\state{ETSI Telecomunicación, Avenida Complutense 30, Madrid 28040}, \country{Spain}}}

\address[2]{\orgdiv{Center for Computational Simulation}, \orgname{Universidad Politécnica de Madrid}, \orgaddress{\state{Campus de Montegancedo UPM, Boadilla del Monte 28660-Madrid}, \country{Spain}}}

\address[3]{\orgdiv{DACYA}, \orgname{Universidad Complutense de Madrid}, \orgaddress{\state{Facultad de Informática UCM, Madrid 28040}, \country{Spain}}}

\corres{*Patricia Arroba, ETSI Telecomunicación, Avenida Complutense 30, Madrid 28040, Spain. \email{parroba@die.upm.es}}

\abstract[Abstract]{
Data centers handle impressive high figures in terms of energy consumption, and
the growing popularity of Cloud applications is intensifying their
computational demand.  Moreover, the cooling needed to keep the servers within
reliable thermal operating conditions also has an impact on the thermal
distribution of the data room, thus affecting to servers' power leakage.
Optimizing the energy consumption of these infrastructures is a
major challenge to place data centers on a more scalable scenario.
Thus, understanding the relationship between power,
temperature, consolidation and performance is crucial to enable an
energy-efficient management at the data center level.
In this research, we
propose novel power and thermal-aware strategies and models to provide joint cooling and computing optimizations
from a local perspective based on the global energy consumption of
metaheuristic-based optimizations.
Our results show that the combined awareness
from both metaheuristic and best fit decreasing algorithms allow us to describe the global
energy into faster and lighter optimization strategies that may be used
during runtime. This approach allows us to improve the energy efficiency of the
data center, considering both computing and cooling
infrastructures, in up to a 21.74\% while maintaining quality of service.
}


\maketitle
\section{Introduction}
\label{sec:introTemp}
Nowadays, data centers consume about the 2\% of the worldwide energy
production~\cite{FP7GreenDataNet}, originating more than 43 million
tons of $CO_2$ per year~\cite{thermal:koomey2011}.  Also, the
proliferation of urban data centers is responsible for the increasing
power demand of up to 70\% in metropolitan areas, where the power
density is becoming too high for the power grid~\cite{451Research}.
In two years, the 95\% of the urban data centers will experience
partial or total outages, incurring in annual costs of about US\$2
million per infrastructure. The 28\% of these service outages are
expected to appear due to exceeding the maximum capacity of the
grid~\cite{PonemonReport}.
The advantages of Cloud computing lie in the usage of a technological
infrastructure that allows high degrees of automation, consolidation and
virtualization, which results in a more efficient management of the resources
of a data center. 
Cloud computing, in the context of data centers, has been proposed as
a mechanism for minimizing environmental impact.  Virtualization and
consolidation increase hardware utilization (of up to
$80\%$~\cite{KatzTechTitans}) thus improving resource
efficiency. Moreover, a cloud usually consists of distributed
resources dynamically provisioned as services to the users, so it is
flexible enough to find matches between different parameters to reach
performance optimizations.
Gartner expectations predict that by 2020,
Cloud adoption strategies would impact on more than the 50\% of the 
IT (Information Technology)
outsourcing deals in an effort to cost optimize the infrastructures
that use from 10 to 100 times more power than typical office
buildings~\cite{Scheihing:CreatingEnergyEfficient:07} even consuming
as much electricity as a city~\cite{Markoff:Intel:02}.

The main contributors to the energy consumption in a data center are:
(i)~the IT resources, which consist of servers and other IT
equipment, and (ii)~the cooling infrastructure needed to ensure that
the IT operates within a safe range of temperatures, ensuring
reliability. The remaining 10\% comes from (iii)~the power consumption
that comes from lightning, generators, uninterruptible power supply systems and
power distribution units~\cite{thermal:koomey2011}.
The IT power in the data center is dominated by the power consumption
of the enterprise servers, representing up to 60\% of the overall data
center consumption.  The power usage of an enterprise server can be
divided into dynamic and static contributions.  
On the other hand, the cooling infrastructure is one of the major
contributors to the overall data center power budget, representing
around $40\%$ of the total power consumed by the entire
facility~\cite{Hamilton_cooperativeexpendable}. This is the main
reason why recent research aim to achieve new thermal-aware techniques
to optimize the temperature distribution in the facility, thus
minimizing the cooling costs.  
Controlling the set point temperature of cooling systems in data
centers is still to be clearly defined and it represents a key
challenge from the energy perspective. This value is often chosen
based on conservative suggestions provided by the manufacturers of the
equipment and it is calculated for the worst case scenario resulting
in overcooled facilities.  Increasing the temperature results in
savings in cooling consumption, so a careful management can be devised
ensuring a safe temperature range for IT resources. 
From the application-framework viewpoint, Cloud workloads present
additional restrictions as 24/7 availability, and SLA (Service Level Agreement) constraints
among others. In this computing paradigm, workloads hardly use the
100\% of the CPU (Central processing unit) resources, and their runtime is strongly
constrained by contracts between Cloud providers and clients. These
restrictions have to be taken into account when minimizing the energy
consumption as they impose additional boundaries to the optimization
strategies.

Just for an average $100kW$ data center, a $7\%$
in annual savings represents around US \$5 million per
year~\cite{Chen_Katz_2009}.
These power and thermal situations have encouraged the challenges in
the data center scope to be extended from performance, which used to
be the main target, to energy-efficiency.  This context draws the
priority of stimulating researchers to develop sustainability policies
at the data center level, in order to achieve a social sense of
responsibility, while minimizing environmental impact.
However, current
approaches do not incorporate the impact of leakage consumption, found
at the technology level, when controlling the cooling set point
temperature at the data center level.  This power contribution, which
grows for increasing temperatures, is neither considered when
optimizing allocation strategies in terms of energy under highly
variable workloads, typically found in Cloud environments.
Thus, using power and thermal models that incorporate this information
at different abstraction levels, which impacts on the power
contributors, helps us to find the relationships required to devise
a global minimization searching for the optimum that combines power and thermal-aware strategies for joint IT and cooling infrastructures.  For this purpose, we evaluate the use of
metaheuristic-based algorithms that provide a global minimization.  In
our work, we will focus on the combination of strategies that are
aware of both IT and cooling consumption, also taking into account the
thermal impact and the resource variability.

This work is intended to offer novel optimization
strategies that take into account the contributions to power of
non-traditional parameters such as temperature and frequency among
others.  Our research is based on fast and accurate models that are
aware of the relationships with power of these parameters, allowing us
to combine both energy and thermal-aware strategies.  Our work makes
the following \textbf{key contributions}: 1) a set of single-objective and
multi-objective BFD (Best Fit Decreasing)-based policies that optimize the energy
consumption of the data center considering both IT and cooling
parameters; 2) a metaheuristic-based optimization policy that relies
in a simulated annealing algorithm to optimize the energy consumption
of both IT and cooling infrastructures; 3) a novel strategy to infer a
local minimization based on modeling the improvements provided by the
simulated annealing optimization;
4) two thermal models that accurately describe the behavior of the CPU
and the memory devices, resulting in an average temperature estimation error of
0.85\% and 0.50\% respectively; and 5) a cooling strategy based on the
estimated temperature of devices due to VM (Virtual Machine) allocation.
The remainder of this paper is organized as follows:
Section~\ref{relWork} gives further information on the related work on
this topic.  Our proposed optimization framework and VM consolidation
algorithms are provided in Sections~\ref{optparad} and~\ref{sec:algorithmDescription}
.  Section~\ref{sec:models} and
Section~\ref{sec:cooling} explain the modeling process for our
metaheuristic-based VM consolidation policy and our cooling strategy
respectively.  Section~\ref{sec:resultsTemp} describes profusely the
experimental results.  Finally, in Section~\ref{sec:conclusionsTemp}
the main conclusions are drawn.

\section{Related Work}
\label{relWork}
Due to the impact of energy-efficient optimizations in an environment
that handles so impressive high figures as data centers, many
researchers have been motivated to focus their academic work on
obtaining solutions for reducing consumption.  
In this section, we present
different approaches of the state-of-the-art based on heuristics and metaheuristics that are aware of energy contributions.

\subsection{Reduce IT power consumption}
Power minimization, resource utilization and Quality of Service (QoS) are the
main targets in energy-efficient strategies for Cloud data
centers~\cite{SUN201592},~\cite{USMANI2016491},~\cite{7152482}.  
Different heuristic algorithms have been used to
solve the VM placement as a bin packing problem.  First fit decreasing 
and Best Fit Decreasing (BFD) policies have been used to minimize the IT energy
consumption from a local
perspective~\cite{Chowdhury2015},~\cite{Beloglazov:2012:OOD:2349876.2349877},~\cite{minPit2014},~\cite{minPit2017}.  These approaches
mainly consider the optimization of a single-objective scenario under certain
QoS constraints.
On the other hand, metaheuristics are also applied to manage the allocation of
each VM in \textit{vmList} will be allocated in the server provided by the best
solution.
VMs in Cloud computing~\cite{7904293}.  Wu et al.~\cite{Wu:2012:EVM:2426936.2426976}
proposes the use of a single-objective GA (Genetic Algorithm) to optimize IT energy consumption.
These metaheuristic-based approaches can be used to solve the problem from a
global perspective (considering the consumption of the whole IT
infrastructure), also targeting more than one optimization objectives. Ye et
al.~\cite{7997707} propose a KnEA-based evolutionary algorithm considering, not
only energy consumption and resource utilization, but also load balance and
robustness objectives simultaneously.  

\subsection{Joint Thermal and Power-Aware considerations}
Currently, data centers save lots of energy by providing efficient
cooling mechanisms.  Hot-spots throughout the facilities are the main
drawback according to system failures, and due to this factor, some
data centers maintain very low room temperatures of up to
13$^{\circ}C$~\cite{KatzTechTitans}. Most recently, data centers are
turning towards new efficient cooling systems that make higher
temperatures possible, around 24$^{\circ}C$ or even 27$^{\circ}C$.
The increasing power density of new technologies has resulted in the
incapacity of processors to operate at maximum design frequency, while
transistors have become extremely susceptible to errors causing system
failures~\cite{Mitra:2005:RSD:1048711.1048754}. Moreover, adding to
this issue that system errors increase exponentially with temperature,
new techniques that minimize both effects are required.
To avoid these thermal-related issues while further reducing
energy, researchers are mainly focusing on considering the power consumption of
servers' fans, the power leakages due to temperature and the power due to air
conditioning units.
\subsubsection{Servers fan power consumption}
Ayoub et al.~\cite{6169035} and Chan et
al.~\cite{Chan:2012:FSD:2333660.2333753} propose energy-aware heuristics
considering joint fan control and scheduling mechanisms to reduce consumption
of a server.  Their work propose a state machine to simultaneously minimize
several objectives. They provide models to estimate temperature that use
electrical analogies to represent the thermal behavior of the components.
\subsubsection{Power Leakage due to temperature} In current electronic systems,
in which integration technology scales below 100nm, static power consumption
represents about 30-50\%~\cite{narendra2010leakage}, of the total power under
nominal conditions.  The impact of temperature, which presents an exponential
dependency on the leakage currents~\cite{rabaey2009low}, aggravates this
situation increasing power consumption.  Some Cloud computing solutions have
taken into account the dependence of power consumption with temperature, due to
both fan speed and induced leakage currents.  In our previous
work~\cite{Arroba:EProcedia:2014}, we present a power model that includes both
contributions, which we used to provide energy savings together with a
DVFS-aware allocation strategy~\cite{arroba2017dynamic}.  In the present
research, we aim to improve energy efficiency by extending this VM allocation
policy, also incorporating thermal-awareness. Li et al.~\cite{6008602} also
provide an heuristic that minimizes power consumption, using a model that
depends on fan power and leakage.
\subsubsection{Power of Air Conditioning Units}
The power budget to cool down data center infrastructures
represents about the 40\% of the total
budget~\cite{Hamilton_cooperativeexpendable}.  By combining energy-aware
strategies in both computing and cooling infrastructures, potential savings of
about $54\%$~\cite{Virt_whitepaper} are estimated.  Thus, researchers are
including the effects of temperature in the data room among their optimization
objectives. 
On its own, virtualization has the potential of minimizing the hot-spot issue
by migrating VMs.  Migration policies allow to distribute the workload during
run-time without stopping task execution. Primas et
al.~\cite{8241106} present an heuristic to perform energy-efficient scheduling
considering the temperature of the data room.  Xu et al.~\cite{5724828} present
a GA-based multi-objective VM placement algorithm that targets power
consumption, resource wastage and thermal dissipation costs.  In both research
works, the energy consumption of the cooling infrastructure is not calculated
but they assume a significant reduction due to reducing hot spots using load balancing.

On the other hand, there are research works in which the power
of air conditioning units is minimized explicitly together with the IT power.
Abbasi et al.~\cite{Abbasi:2010:TAS:1851476.1851493} propose heuristic
algorithms to address this problem.  Their work presents the data center as a
distributed cyber physical system in which the energy of both IT and
cooling is the minimization objective.  Li et al.~\cite{7888576} use a greedy
scheduling algorithm, considering the temperature distribution airflow to
minimize the energy consumption of IT and cooling jointly.
Current research in the area of joint energy and thermal aware
strategies consider fan power, leakage due to temperature or cooling units'
power, but do not take into account the combined effects between them.  Our
work proposes the global minimization of the energy of the entire facility (IT
and cooling infrastructures) considering all these thermal effects and their
impact on power consumption in a proactive way.

\section{Optimization paradigm}
\label{optparad}
\begin{figure} [b]
  \centering
  \includegraphics[width=0.4\columnwidth]{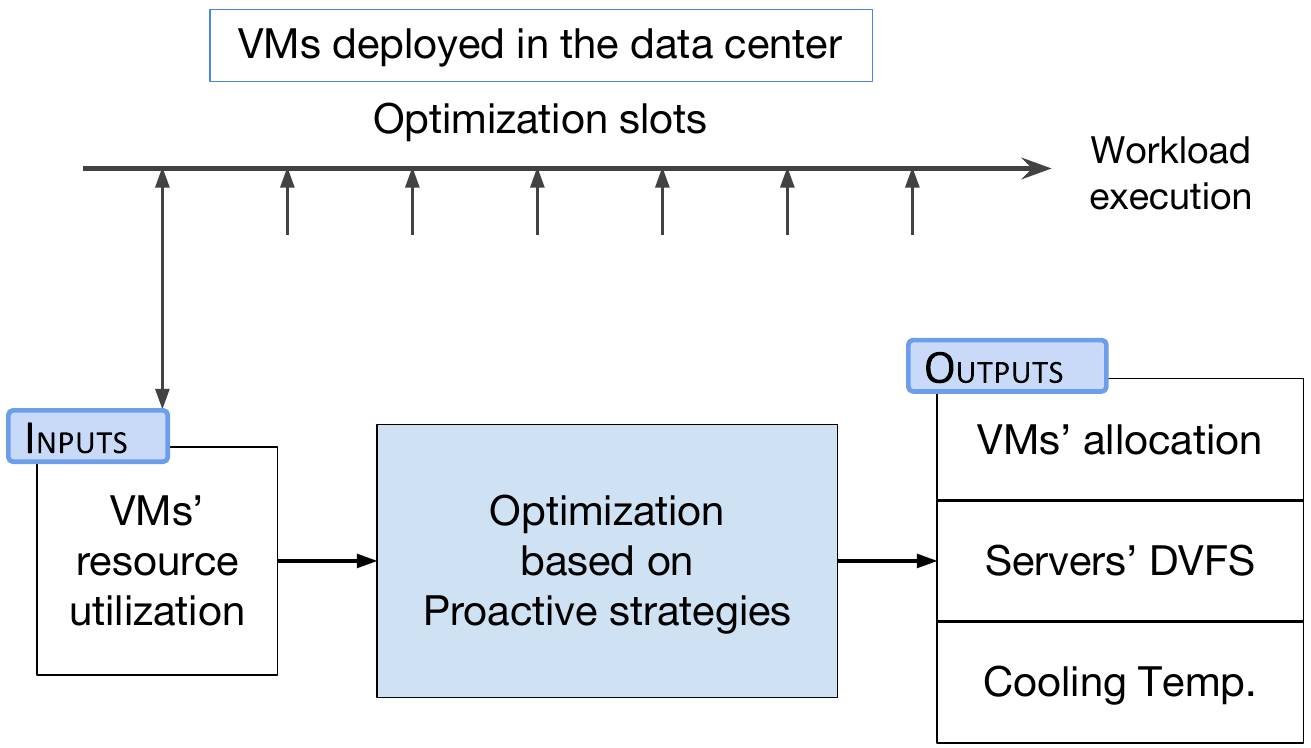}
  \caption{Overview of the inputs and outputs of the proposed optimization framework.}
  \label{fig:inoutframework}
\end{figure}

The new simulation and optimization framework presented in this research considers the
energy of a Cloud data center from a global and proactive perspective.
So, our proposed optimization algorithms are aware of the evolution of
the global energy demand, the thermal behavior of the room and the
workload considerations at the different data center subsystems during
runtime.
This section explains our hypothesis on how to provide a feasible
and proactive solution to approach the issues that have been
motivated.  The scenario chosen for the development of this research,
is a Cloud optimization framework that can be seen in
Figure~\ref{fig:inoutframework}.  The applications require constant
monitoring of their computational demand in order to capture their
variability during runtime and to perform VM migrations when needed in
order to avoid QoS degradation.

\begin{figure} [htb]
  \centering
  \includegraphics[width=0.6\columnwidth]{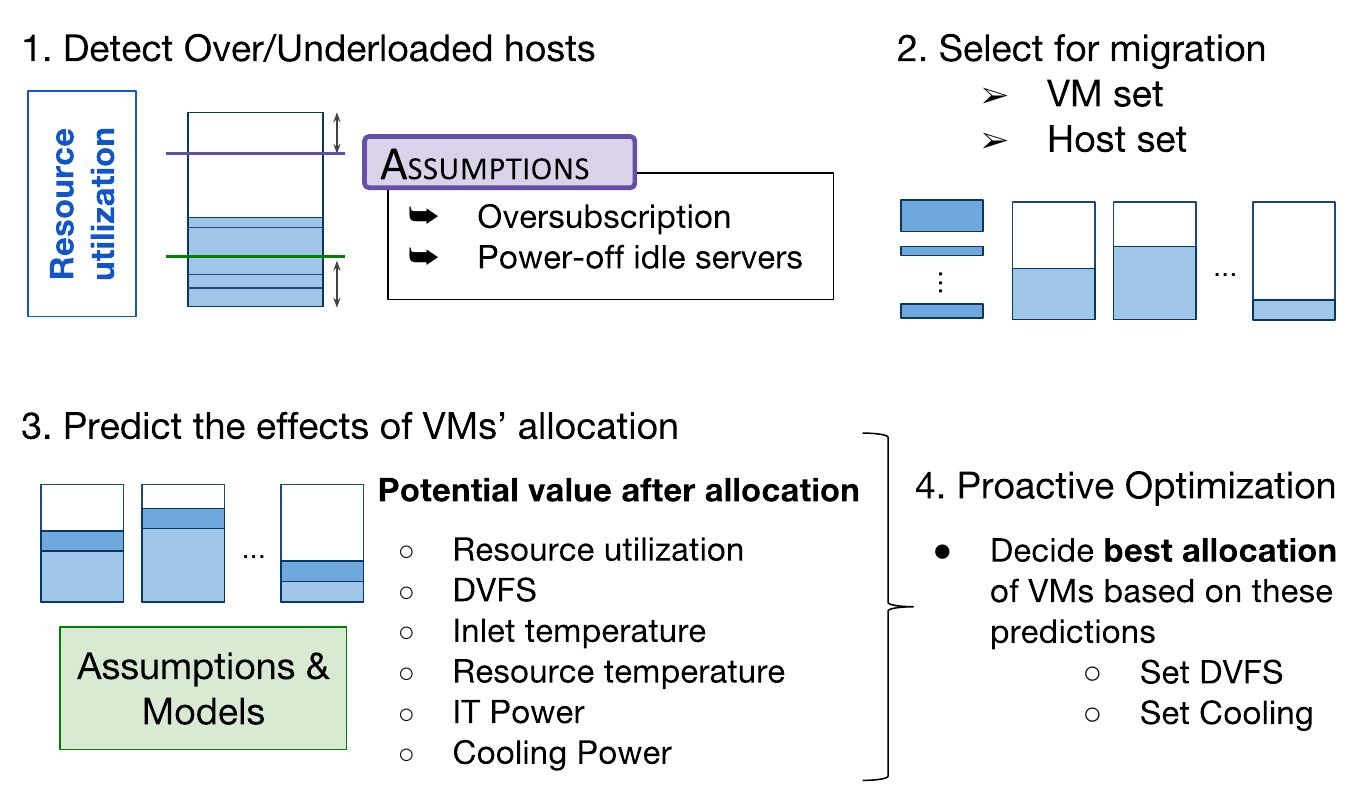}
\setlength{\belowcaptionskip}{-20pt}
  \caption{Overview of the optimization diagram for the proposed framework.}
  \label{fig:optdiag}
\end{figure}

As during this research work we did not have access to an operative
Cloud data center, we leverage real traces publicly released by Cloud
providers to simulate the operation of the infrastructure. These
traces consist of periodic resource usage reports that provide
specific information as CPU demand percentages and memory and disk
usage and provisioning values for all VMs.  These traces are the only
input for our optimization framework, based on proactive
strategies. Finally, for each optimization slot, we obtain the
allocation for the VMs in the system, as well as the servers'
DVFS configuration and the cooling set point temperature.
In Figure~\ref{fig:optdiag}, we present our proposed optimization
based on proactive strategies more in detail. For each optimization
slot, in which we have input traces, we detect overloaded hosts for
the current placement of VMs that are already deployed on the system,
where oversubscription is allowed.  Overloaded hosts are more likely
to suffer from performance degradation, so some VMs have to be
migrated from them to other hosts.
Based on this information, the consolidation algorithm selects: 1) the
set of VMs that have to be migrated from the overloaded physical
machines and 2) the set of servers that are candidates to host these
VMs.  Then, the models help to predict the effects of potential
allocations for the set of VMs. These models are needed to provide
values of both the parameters that are observed in the infrastructure
(temperature and power of the different resources) and the control
variables (VM placement, DVFS and cooling set point temperature).
Finally, the proactive optimization algorithm decides the best
allocation of VMs based on these predictions. After this first
iteration, if underloaded hosts are found, this optimization process
is repeated in order to power off idle servers if possible.
We assume this data center may be homogeneous in terms of IT
equipment.  Live migration, oversubscription of CPU, DVFS and
automatic scaling of the active servers' set are enabled.  We assume a
traditional hot-cold aisle data center layout with CRAC (Computer Room Air Conditioning)-based cooling.
In particular, at the data center level we consider a raised-floor
air-cooled infrastructure where cold air is supplied via the floor
plenum and extracted in the ceiling.

\section{Description of VM Allocation Strategies} 
\label{sec:algorithmDescription}

The different policies presented in this section are based on the
knowledge achieved from our power model presented in our previous
research~\cite{Arroba:EProcedia:2014} for a Fujitsu RX300 S6 server, that is
further explained in Section~\ref{sec:resultsTemp} and can be seen in
Equation~\ref{eq:powerModel}.
In this section we provide a taxonomy of candidate optimization
algorithms that take into account IT and cooling power of the data
center infrastructure under energy and thermal considerations. The
mathematical description of these objectives will allow the later
optimization by the development of an optimization algorithm.

\subsection{Single-Objective BFD-based Allocation Policies}
\label{SOlabel}
These Single-Objective (SO) policies optimize the consolidation of a set of VMs
($vmList$) using the BFD approach. First, VMs are sorted in decreasing order of
CPU utilization and then, they are allocated on the set of available hosts
($hostList$) according to the minimization of an optimization objective that is
calculated by the function $SOvalue()$ as can be seen in
Algorithm~\ref{alg:SOAlgorithm}. This function computes the single objective
according to the model selected from those presented in the following
Subsections~\ref{SO1}-~\ref{SOend}. $bestPlacement$ and $bestHost$ are the best
placement value for each iteration and the best host to allocate the VM
respectively. 

\renewcommand{\algorithmicrequire}{\textbf{Input:}}
\renewcommand{\algorithmicensure}{\textbf{Output:}}
\renewcommand{\algorithmicforall}{\textbf{foreach}}
\begin{algorithm} [htb]
\caption{SO Placement Policy}
\label{alg:SOAlgorithm}
\begin{algorithmic}[1] 
\Require {\normalsize hostList}, {\normalsize vmList}
\Ensure {\normalsize SOPlacement} of VMs
\State {\normalsize vmList}.\texttt{sortDecreasingUtilization()}
\ForAll {{\normalsize vm} \textit{in} {\normalsize vmList}}
        \State {{\normalsize bestPlacement} $\gets$ MAX}
        \State {{\normalsize bestHost} $\gets$ NULL}
        \ForAll {{\normalsize host} \textit{in} {\normalsize hostList}}
                \If {{\normalsize host} \textit{has enough resources for} {\normalsize vm}}
                        \State {{\normalsize placement} $\gets$ {\normalsize SOvalue()}}
                        \If {{\normalsize placement} $<$ {\normalsize bestPlacement}}
                                \State {{\normalsize bestHost} $\gets$ {\normalsize host}}
                                \State {{\normalsize bestPlacement} $\gets$ {\normalsize placement}}
                        \EndIf
                \EndIf
        \EndFor
        \If {{\normalsize bestHost} $\neq$ NULL}
                \State {{\normalsize SOPlacement}.\texttt{add}({\normalsize vm}, {\normalsize bestHost})}
        \EndIf
\EndFor
\State \textbf{Return:} {\normalsize SOPlacement}
\end{algorithmic}
\end{algorithm}

Then, each VM in $vmList$ will be allocated in a server that belongs
to the list of hosts that are not overutilized ($hostList$) and have
enough resources to host it. The VM is allocated in the host that has
the lowest $placement$ value. The output of this algorithm is the
placement ($SOPlacement$) of the VMs that have to be mapped according
to the MAD-MMT (Median Absolute Deviation - Minimum Migration Time) detection and VM selection policy as in the research
presented by Beloglazov et
al.~\cite{Beloglazov:2012:OOD:2349876.2349877}.  In this subsection we
present the different SO objectives proposed in this research.

\begin{figure} [tb]
\centering
\includegraphics[width=0.5\columnwidth]{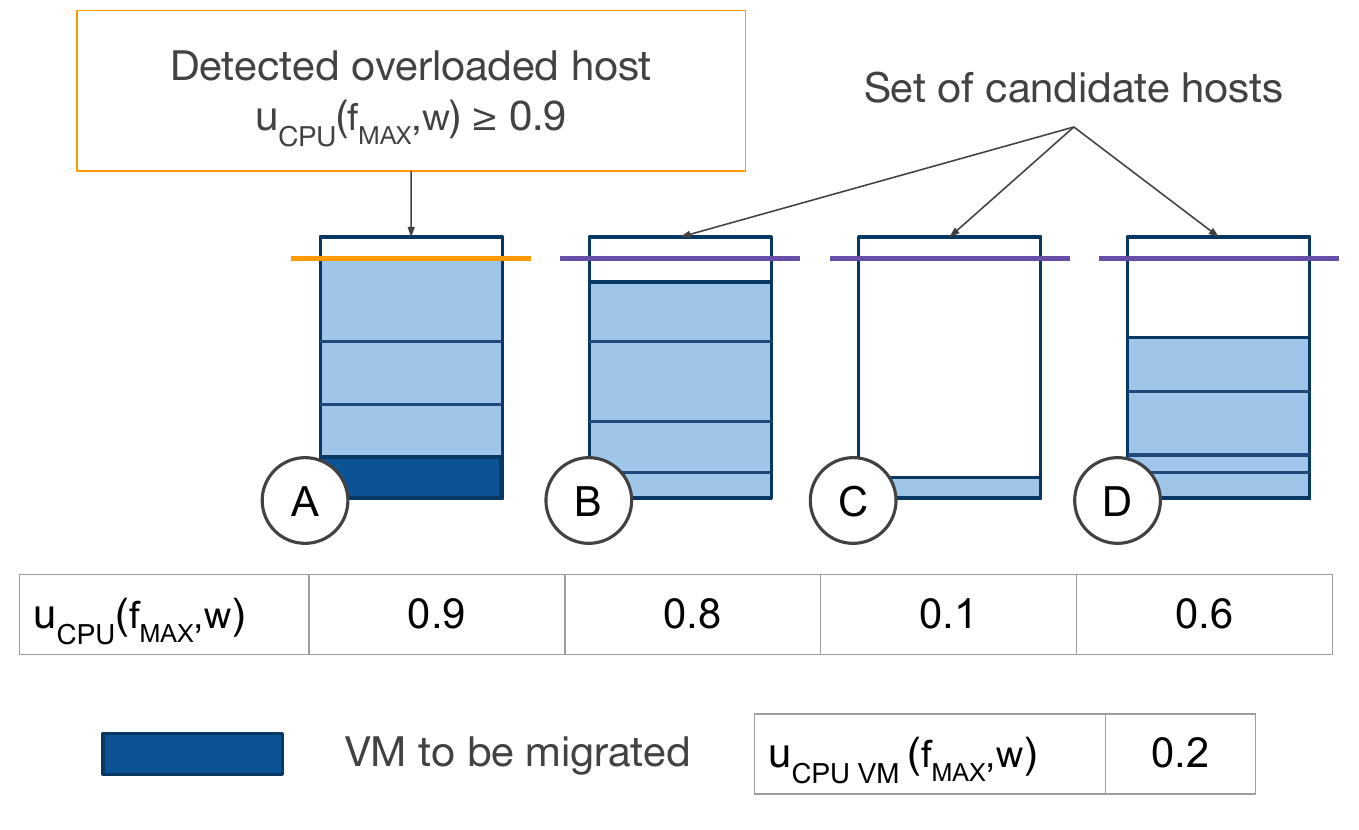}
\setlength{\belowcaptionskip}{-20pt}
\caption{Case of use of VM allocation algorithm. Initial status.}
\label{fig:caseOfUseInitialStatus}
\end{figure}

Additionally, we provide the following case of use in
Figure~\ref{fig:caseOfUseInitialStatus} to explain the proposed concepts.  In
this example, the system detects an overloaded situation in host A, and the
algorithm will reallocate the VM that may be migrated to one of the candidate
hosts (B-D). The $u_{CPU}$ values provide the current CPU utilizations of the servers and the VM. Figure~\ref{fig:caseOfUse}, located at the end of the
subsection, compiles the key monitoring parameters needed by the set of policies. The values after the allocation are
estimated using the models proposed in this paper, which can be seen in
Subsection~\ref{powAndTherModels}.
\subsubsection{$SO_{1}$: $min\{{\Delta} P_{Host}\}$}
\label{SO1}
This policy minimizes the increment of power consumption when a VM is
allocated in a host. The algorithm consolidates the VM in the server
whose power consumption has the lowest increase in terms of power. The
increment is calculated as the difference between the power after and
before the allocation of the incoming VM.  
In the case of use presented in this subsection, the CPU
utilization values after allocation of servers B, C and D are 1, 0.3 and 0.8
respectively.  For this example, a host is considered overloaded if its CPU
utilization is equal to or higher than 0.9, so host B would not be a valid
candidate to host the VM.  According to Figure~\ref{fig:caseOfUse}, host power
increment after the allocation of servers C and D is 10~W and 20~W
respectively.  So, the $SO_{1}$ policy chooses server C as the best candidate
to host the VM, as it provides a higher CPU utilization after the allocation.
This approach has been proposed by Beloglazov et al.~\cite{Beloglazov:2012:OOD:2349876.2349877} in their PABFD algorithm.
We present this policy in this section, as we include it in our multi-objective approaches in the following sections.

\subsubsection{$SO_{2}$: $min\{P_{Host}\}$}
The consolidation value is the host's power consumption. The algorithm chooses
the server that presents the lowest power consumption when hosting the incoming
VM.  In our case of use the host power values after allocation are 170~W and
200~W respectively for servers C and D. So, the $SO_{2}$ policy chooses server
C as the best candidate to host the VM.  The main objective of this policy is
to support the need of understanding the different contributions to power in a
data center and it is available in the
state-of-the-art~\cite{minPit2014},~\cite{minPit2017}.  The fact is that the
global power consumption can not be reduced by minimizing local power, as the
static contribution increases globally with the number of active hosts.
However, this consolidation technique may be useful in some scenarios in which
power capping is a necessity, enforcing a drastic reduction of server power.

\subsubsection{$SO_{3}$: $min\{1/(u_{cpu}-\Delta freq)\}$}
The consolidation value calculated by this approach has been proposed
by the authors in previous work~\cite{arroba2017dynamic}. Our
proposed policy is not only aware of the utilization of the incoming
VM, but also considers the impact of its allocation in terms of
frequency.  
In our case of use, according to Figure~\ref{fig:caseOfUse}, the frequency
increment after the allocation of servers C and D is 0 and 0.4 respectively, so the allocation values are 1/0.3 and 1/(0.8-0.4).
So, the $SO_{3}$ policy
chooses server D as the best candidate to host the VM.
This approach is interesting from the point of view of
combining both static and dynamic contributions to global power
consumption from a local perspective. The higher the CPU utilization
allowed in servers, the lower the number of active hosts required to
execute the incoming workload, thus reducing global static
consumption.  However, host's dynamic consumption increases with
frequency. As frequency increases with CPU utilization demand, we
propose a compromise between increasing the $u_{cpu}$ after the
allocation and reducing the frequency increment due to the incoming
VM. This equation may be devised as both the $u_{cpu}$ and the
frequency increment range in the same orders of magnitude.

\subsubsection{$SO_{4}$: $min\{T_{mem}\}$}
This consolidation approach aims to minimize the temperature of the
memory as it has been demonstrated to be a key contribution to static
power consumption.  
In our case of use, the memory temperature values after allocation are
38\degree C and 55\degree C respectively for servers C and D. So, the $SO_{4}$ policy
chooses server C as the best candidate to host the VM.
Moreover, this parameter also depends on the inlet
temperature of the server, which impacts on the cooling power of the
data center infrastructure and on the dynamic memory activity.
Cooling down the computing infrastructure is needed to avoid failures
on servers due to temperature, or even the destruction of components
as in the case of thermal cycling and electromigration among others.
Therefore, this policy may be helpful for extremely hot conditions in
the outside, and also when undergoing cooling failures.

\subsubsection{$SO_{5}$: $min\{\Delta freq\}$}
\begin{figure} [htb]
\centering
\includegraphics[width=0.7\columnwidth]{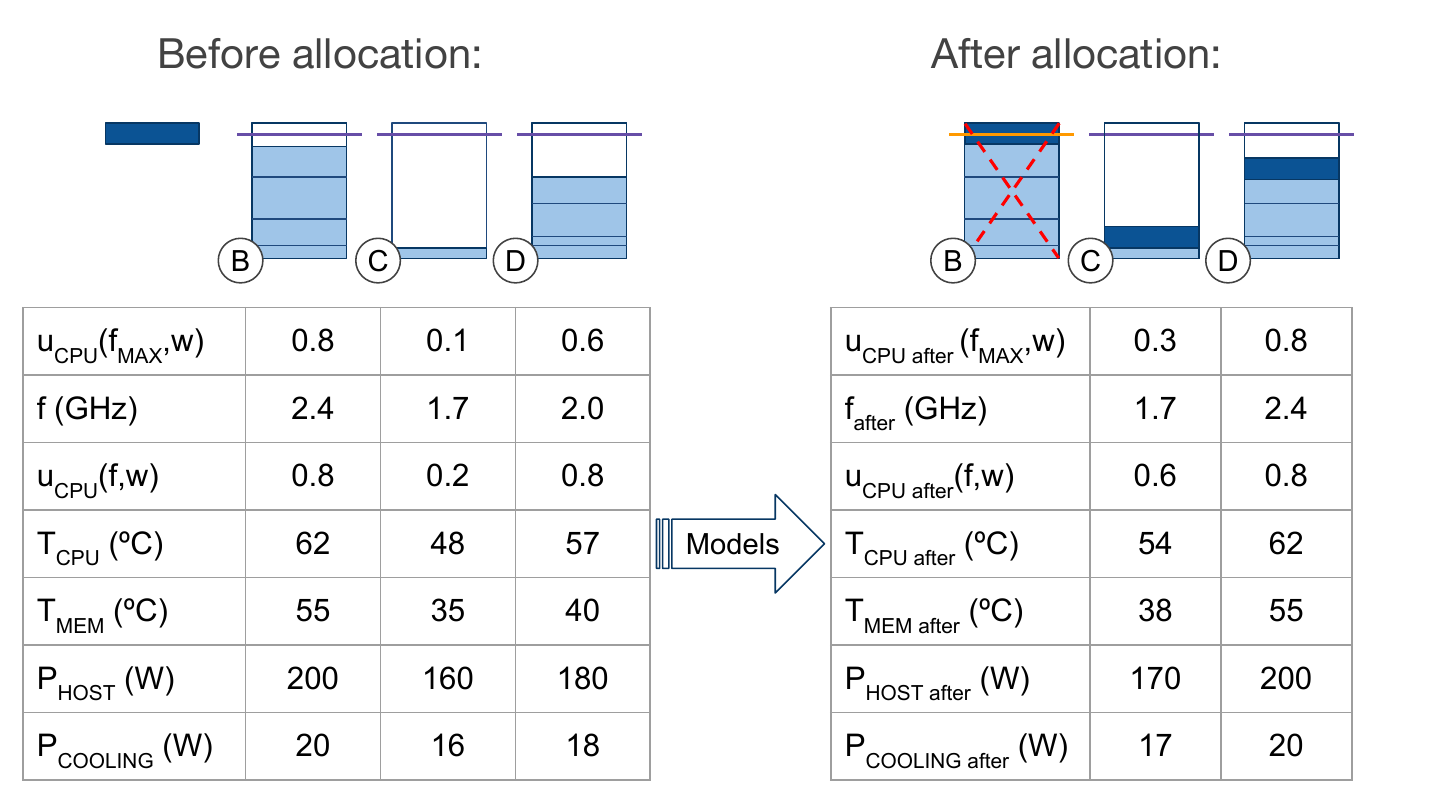}
\setlength{\belowcaptionskip}{-20pt}
\caption{Case of use of VM allocation algorithm. Values before and after VM allocation.}
\label{fig:caseOfUse}
\end{figure}

This algorithm consolidates the VM in the server whose frequency has a
lowest increase. The policy aims to minimize the increment of power
consumption when a VM is allocated in a host. The increment is
calculated as the difference between the frequency after and before
the allocation of the VM.  
In our case of use, according to Figure~\ref{fig:caseOfUse}, the frequency
increment after the allocation of servers C and D is 0 and 0.4 respectively. So, the $SO_{5}$ policy
chooses server C as the best candidate to host the VM.
This consolidation technique aims to
minimize the dynamic contribution to power consumption.

\subsubsection{$SO_{6}$: $min\{1/u_{cpu}\}$}
The consolidation value proposed in this approach maximizes the
overall CPU utilization in the active host set, also constraining the
number of active servers. 
In the case of use presented in this subsection, the CPU
utilization values after allocation of servers C and D are 0.3 and 0.8
respectively.  So, the $SO_{6}$ policy chooses server D as the best
candidate to host the VM.
In all our proposed approaches, the maximum
load that can be allocated in each host is bounded by its available
resources. Also, the VMs are migrated from overloaded servers
according to previous workload variations. However, the possibilities
of violating the SLA are high when using this specific technique.
This is because workload variations in a highly loaded server may
exceed the total resource capacity of the device, degrading the
performance of the applications. This policy may be specially useful
in scenarios with low penalties per SLA violations or if performance
degradation does not have a great impact on economic or contractual
issues.

\subsubsection{$SO_{7}$: $min\{P_{Host}+P_{Cooling}\}$}
The consolidation value is the aggregation of the host's power
consumption and the power dimensioned to cool it down avoiding thermal
issues. The algorithm chooses the server that presents the lowest
power IT and cooling consumption when hosting the incoming VM. 
In our case of use, the host and cooling aggregated power values after allocation are
187~W and 220~W respectively for servers C and D. So, the $SO_{7}$ policy
chooses server C as the best candidate to host the VM.
The
main objective of this policy is to show the relevance of
understanding the thermal contributions to power in a data center.
The overall power consumption can not be reduced by minimizing local
power, as the static IT contribution increases globally with the
number of active hosts and the cooling power depends on IT consumption
as well as on their inlet temperature needed to keep them
safe. However, this consolidation technique may be useful in some
scenarios in which power capping is a necessity in both IT and cooling
infrastructures and when combined with variable cooling techniques.

\subsubsection{$SO_{8}$: $min\{\sum|SO_{X}|\}$} 
\label{SOend}
This approach aims to minimize the total power consumption of the data
center by combining the different parameters presented in the SO
section in a single metric. In order to make the values comparable, we
normalize them in the range [1,2] (according to their maximum and
minimum in each range) so all the inputs take values in the same
orders of magnitude. It is worth to mention that we have performed a
variable standardization for every feature in order to ensure the same
probability of appearance for all the variables. The algorithm
consolidates the VM in the host that minimizes the summation of all
the normalized parameters as seen in Equation~\ref{eq:so8basic}.
After an exhaustive analysis, the best global power results are shown
for the parameter combination presented in Equation~\ref{eq:so8best}.
\begin{eqnarray}
\label{eq:so8basic}
SO_{8} & = & min \{|\Delta P_{Host}| + |P_{Host}| + |1/(u_{cpu}-\Delta freq)|
+ |T_{mem}| + |\Delta freq| + |1/u_{cpu}| + |P_{Host}+P_{Cooling}|\} \\
\label{eq:so8best}
SO_{8} & = & min\{|1/(u_{cpu}-\Delta freq)| + |\Delta freq| +
|1/u_{cpu}|\}
\end{eqnarray}

\subsection{Multi-Objective BFD-based Allocation Policies}
The approaches that we present in this section also aim to minimize
global power consumption from a local perspective. However, these
policies do not consider one single objective to be optimized, but
more than one. Multi-Objective (MO) optimizations try to simultaneously optimize
several contradictory objectives.  This is also useful due to the fact
that, in some cases, the parameters cannot be linearly combined as
they are not comparable without normalization (e.g. their
orders of magnitude are very different).
In all our MO policies, the VMs from \textit{VMlist} are also allocated one
by one in the host that minimizes the consolidation value according to
the given policy.  However, MO techniques offer a multidimensional
space of solutions instead of returning a single value.  For this kind
of problems, single optimal solution does not exist, and some
trade-offs need to be considered.  The number of dimensions is equal
to the number of objectives of the problem. Each objective of our MO
strategy consists of each one of the SO consolidation values presented
in Subsection~\ref{SOlabel}. Hence, to find the solution for the
allocation of a VM in a specific $Host_i$ from \textit{hostList}, we calculate the
consolidation value of all the SO policies in \textit{SOList} (\{$SO_{1}, ..., SO_{7}$\}) using the \textit{SOvalue()} function as can be seen in Algorithm~\ref{alg:MOAlgorithm}. We do not include the single objective $SO_{8}$ in this calculations as its resulting value is a linear combination of the rest of SO objectives. Thus, each solution of the
algorithm has the following multi-objective vector (\textit{hostVector}):
\begin{eqnarray}
solution_{Host_i} = \{\Delta P_{Host_i}, P_{Host_i}, 1/(u_{cpu}-\Delta freq)_i, 
T_{mem_i}, \Delta freq_i, 1/u_{cpu_i}, P_{Host_i} + P_{Cooling_i}\}
\end{eqnarray}

Then, we constrain the set of solutions to provide the ones that are
in the Pareto-optimal Front (POF) \textit{paretoOptimal}.  This optimal
subset provides only those solutions that are non-dominated by others
in the entire feasible search space.
This approach discards solutions that may be the optimum for a SO policy, but
are dominated by other solutions that appear in MO problems.  The \textit{MOvalue()}
function computes the objective according to the model selected from those
presented in the following subsections~\ref{MO1} and~\ref{MO2}.
\textit{bestPlacement} and \textit{bestHost} are the best placement value for each iteration
and the best host to allocate the VM respectively. 
Then, each VM in \textit{vmList} will be allocated in a server that belongs to the
list of hosts whose \textit{hostVector} are non-dominated and have enough resources to
host it.  The VM is allocated in the host that has the lowest \textit{placement}
value.  The output of this algorithm is the placement (\textit{MOPlacement}) of the
VMs that have to be mapped according to the MAD-MMT detection and VM selection
policy.  In this section, we present two MO metrics to decide a solution from
the Pareto-optimal set of solutions.

\begin{algorithm} [htb]
\caption{MO Placement Policy}
\label{alg:MOAlgorithm}
\begin{algorithmic}[1]
\Require {\normalsize hostList}, {\normalsize vmList}, {\normalsize SOList}
\Ensure {\normalsize MOPlacement} of VMs
\State {\normalsize vmList}.\texttt{sortDecreasingUtilization()}
\ForAll {{\normalsize vm} \textit{in} {\normalsize vmList}}
        \State {{\normalsize bestPlacement} $\gets$ MAX}
        \State {{\normalsize bestHost} $\gets$ NULL}
        \ForAll {{\normalsize host} \textit{in} {\normalsize hostList}}
                \If {{\normalsize host} \textit{has enough resources for} {\normalsize vm}}
			\ForAll {{\normalsize SO} \textit{in} {\normalsize SOList}}
                        	\State {{\normalsize hostVector}.\texttt{add}({\normalsize SOvalue()})}
			\EndFor
                        \State {{\normalsize hostVectorSolutions}.\texttt{add}({\normalsize host}, {\normalsize hostVector})}
                \EndIf
        \EndFor
       	\State {{\normalsize paretoOptimal} $\gets$ {\normalsize hostVectorSolutions}.\texttt{getNonDominatedHostVectors()}}
        \ForAll {{\normalsize host} \textit{in} {\normalsize paretoOptimal}}
                \State {{\normalsize placement} $\gets$ {\normalsize MOvalue()}}
                \If {{\normalsize placement} $<$ {\normalsize bestPlacement}}
                        \State {{\normalsize bestHost} $\gets$ {\normalsize host}}
                        \State {{\normalsize bestPlacement} $\gets$ {\normalsize placement}}
                        \EndIf
               \EndFor
        \If {{\normalsize bestHost} $\neq$ NULL}
                \State {{\normalsize MOPlacement}.\texttt{add}({\normalsize vm}, {\normalsize bestHost})}
        \EndIf
\EndFor
\State \textbf{Return:} {\normalsize MOPlacement}
\end{algorithmic}
\end{algorithm}

\subsubsection{$MO_{1}$: $min\{\sum (P_{host} + P_{Cooling})\}$}
\label{MO1}
To allocate each VM, we consider the solution from the Pareto-optimal
set that provides the lowest global IT and cooling power. Thus, for
every solution in the POF, the algorithm calculates the $MO_1$
consolidation value as the power consumed by the data center
considering the VM placement. Then, the VM is allocated in the host
that minimizes this consolidation value.

\subsubsection{$MO_{2}$: $min\{|d(solution_{Host_i},o)|\}$}
\label{MO2}
For each solution in the POF, the algorithm calculates the Euclidean
distance from the objective vector $solution_{Host_i}$ to the
origin. Finally, the VM is allocated in the host that minimizes this
distance.

\subsection{Metaheuristic-based Allocation Policies}
In order to compare our algorithms with non-local policies we consider
a different approach.  Local search methods usually fall in suboptimal
regions where many solutions are equally fit.  The method proposed in
this section intends to help the solutions to get out from local
minimums finding better solutions. Metaheuristics aim to
optimize the global power consumption by simultaneously allocating all
the VMs in the available host set, instead of in a sequential way as
in our BFD-based algorithms. Thus, these algorithms do not only
consider local power in each server but the entire IT and cooling data
center consumption during the allocation.

\subsubsection{\textit{Simulated Annealing:} $min\{\sum (P_{Host} + P_{Cooling})\}$}
\label{SAsect}
\begin{figure} [b]
\centering
\includegraphics[width=0.6\columnwidth]{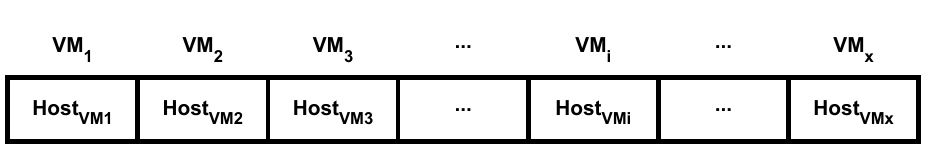}
\setlength{\belowcaptionskip}{-20pt}
\caption{Solution scheme for the SA algorithm.}
\label{fig:SAchrom}
\end{figure}

The Simulated Annealing (SA)~\cite{Kirkpatrick83optimizationby} is a metaheuristic based on the physical
annealing procedure used in metallurgy to reduce manufacturing
defects. The material is heated and then it is cooled down slowly in a
controlled way, so the size of its crystals increases and, in
consequence, this minimizes the energy of the system. SA is used
for solving problems in a large search space, both unconstrained and
bound-constrained, approximating the global optimum of a given
function.  This algorithm performs well for problems in which an
acceptable local optimum is a satisfactory solution, and it is often
used when the search space is discrete.
Our version of SA proposed in this research minimizes the power consumption of
the data center after the consolidation of the VM set along the infrastructure.
We provide the structure of the SA allocation problem in Figure~\ref{fig:SAchrom}, where each
($VM_i$) is hosted in $Host_{VM_i}$. The size of the solution is the size of
the list of VMs that have to be allocated in the system.  The allocation is
performed for the solution that minimizes the global data center power,
considering both IT and cooling contributions.  The objective function
\textit{finalObjective} aggregates a power objective (\textit{objectivePower})
and a feasibility constraint (\textit{feasibilityConstraint}) according to
Equation~\ref{eq:objsSA}.
\begin{eqnarray}
\label{eq:objsSA}
finalObjective = objectivePower * (1 + feasibilityConstraint)
\end{eqnarray}
where \textit{objectivePower} is the summation of the
power consumption of all the hosts and cooling for each solution. This function
also incorporates penalties in the solution evaluation for those servers that
are overutilized after the consolidation process in terms of CPU utilization,
RAM memory or I/O Bandwidth.  \textit{feasibilityConstraint} is zero if the
solution proposes a VM allocation for the entire set of VMs that meets all the
requirements for all the hosts. Otherwise, this term is the summation of
the excess of CPU, RAM and I/O Bandwidth for the entire host set and is several orders of magnitude higher than the objectivePower term. This approach
reduces the probability of obtaining a solution that is not feasible in our
scenario.
SA algorithms are usually initialized with a random solution. Due to the large space of solutions of our problem, this approach
was unable to provide feasible solutions. So, we improved our SA by
initializing the first iteration with the best solution found by the SO set of
algorithms ($SO_1$ to $SO_8$). This initialization ensures that the SA finds a
feasible solution that is, at least, as good as the one provided by the best SO
in each optimization slot. Steps 1 and 2 of our Metaheuristic Placement Policy shown in Algorithm~\ref{alg:SAAlgorithm} perform this initialization.

Then, the description of the problem (\textit{problemSA}) is provided by this initial solution, together with the VMs in
\textit{vmList} that have to be placed simultaneously within the \textit{hostList} set.
Step 4 in Algorithm~\ref{alg:SAAlgorithm} provides the best solution (\textit{bestSolution}) obtained by the SA.
Finally, the algorithm stores the placement of all the VMs in the hosts provided by \textit{bestSolution} in the output \textit{MetaheuristicPlacement}.
As we would like to compare our work with a metaheuristic-based global algorithm, we will use this algorithm as a baseline for the rest of local BFD-based policies.
\begin{algorithm} [tb]
\caption{Metaheuristic Placement Policy}
\label{alg:SAAlgorithm}
\begin{algorithmic}[1]
\Require {\normalsize hostList}, {\normalsize vmList}, {\normalsize bestSolutionSO}, {\normalsize bestPlacementSO}
\Ensure {\normalsize MetaheuristicPlacement} of VMs
\State {{\normalsize bestPlacement} $\gets$ bestPlacementSO}
\State {{\normalsize bestObjective} $\gets$ bestSolutionSO}
\State {{\normalsize problemSA} $\gets$ {\normalsize \texttt{problem}(vmList, hostList, bestPlacement)}}
\State {{\normalsize bestSolution} $\gets$ {\normalsize \texttt{SimulatedAnnealing}(problemSA)}}
\ForAll {{\normalsize vm} \textit{in} {\normalsize bestSolution}}
        \State {{\normalsize bestHost} $\gets$ {\normalsize {vm.\texttt{getHost()}}}}
        \State {{\normalsize MetaheuristicPlacement}.\texttt{add}({\normalsize vm}, {\normalsize bestHost})}
        
\EndFor
\State \textbf{Return:} {\normalsize MetaheuristicPlacement}
\end{algorithmic}
\end{algorithm}

\subsection{Metaheuristic-based SO Allocation Policies}
Metaheuristics, as SA, help us to achieve a global minimization searching for the optimum, but,
however, may take a higher simulation time when compared to local
policies.  For very complex scenarios, the time required by the SA to
find the solution may exceed the time fixed for the optimization slot,
making it unfeasible to use this metaheuristic during runtime.  On the
other hand, local policies (both SO and MO), provide fast
optimizations, but their solutions may fall into suboptimal regions as
they rely on local information.
In this subsection we present a novel strategy to derive a global
minimization from a local perspective based on modeling the global
energy consumption of metaheuristic-based optimizations.  Our approach
is used to model a new SO policy that combines the different SO
consolidation metrics presented in Subsection~\ref{SOlabel} in order
to find a local policy that outperforms their single outcomes. By
using this modeling technique, we aim to find a local, fast and light
consolidation algorithm that is aware of the global relationships
between the contributions to energy, not only during the allocation,
but also taking into account further VM migrations.

First, we monitor the global energy values obtained during the
simulation of the SA optimization. Under the same context, we run the
SO simulation and monitor the global energy values and the local
consolidation values for the same workload, so the optimization slots
are the same.  Using the optimization slots, we align the data from
both sources and obtain only the optimization slots in which the SA
outperforms the SO solutions in terms of global energy. This is
necessary to model the improvement provided by the joint IT and cooling energy
minimization, because the SA is initialized to the best SO solution
(explained in Subsection~\ref{SAsect}) and in some cases the
metaheuristic is not able to find a better solution.  Based on these
values, we model the global energy of the SA using the SO
consolidation and energy values as in Equation~\ref{eq:funSOSA}.
Then, we use this function to provide a $SO_{SA}$ local consolidation
value for optimizing the system as done before for the regular SO
policies. So, we use $SO_{SA}$ as the other SO policies, allocating
the VMs of the set, one by one, in the host that offers a lowest
consolidation value (as detailed in Algorithm~\ref{alg:SOAlgorithm}).
Further details regarding the implementation of this strategy are
provided in Section~\ref{sec:models}.
\begin{eqnarray}
\label{eq:funSOSA}
SO_{SA} = E_{SA} = f(SO_1, E_{SO_1}, SO_2, E_{SO_2}, ..., SO_8, E_{SO_8})
\end{eqnarray}

\subsection{BFD-based SO Dynamic selection Allocation Policy}
\begin{figure} [htb]
\centering
\includegraphics[width=0.6\columnwidth]{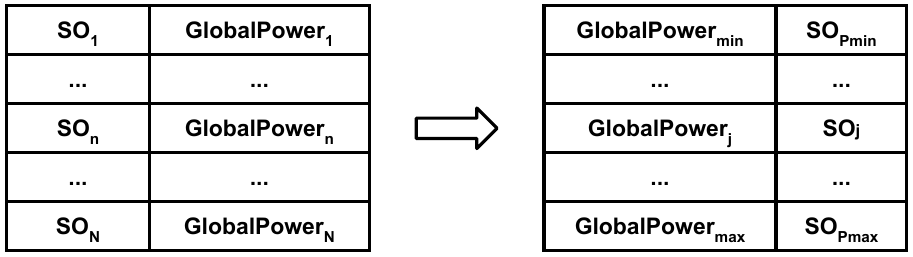}
\setlength{\belowcaptionskip}{-20pt}
\caption{Dynamic selection of the best SO policy.}
\label{fig:DynSO}
\end{figure}

The SO strategies provided in this research offer different allocation
policies from a local perspective.  However, the Cloud computing
context presented in this research shows a very high variability
during runtime in terms of resource demand.  According to the SO
approaches, in each optimization slot, the system only has the
information of a specific instantaneous state and local policies are
not aware of the variability during runtime.
The approach that we present in this section also aims to minimize
global power consumption from a local perspective but is intended to
adapt the system optimization to the dynamism of the Cloud
environment. Therefore, this policy does not consider only one
consolidation value, but the complete set of values offered by all the
SO approaches in this research. The SO Dynamic Selection approach
(\textit{DynSO}) allocates the set of VMs using the SO policy that minimizes
the overall IT power in each time slot.
\begin{eqnarray}
SO \in \{SO_1, ... , SO_{n}, ... , SO_{N}\} 
\end{eqnarray}
We define $SOList$ as $\{SO_{1}, ..., SO_{8}, SO_{SA}\}$, incorporating all the SO policies provided in this research.
Algorithm~\ref{alg:DynSOAlgorithm}
presents the implementation for this allocation policy.
In the first iteration, the algorithm evaluates the final global power consumption
($GlobalPower_n$) provided by the allocation of the entire set of VMs in \textit{VMList}, as in Algorithm~\ref{alg:SOAlgorithm},
using the consolidation value of the first SO in \textit{SOList} ($SO_{n} = SO_{1}$). 
The obtained placement map (\textit{tentativeSOPlacement}) is used to get the global power provided by this allocation policy (\textit{GlobalPowerSO}), using the function \textit{getSOGlobalPower}.
Then, the algorithm stores the global power, and the SO used in the \textit{PowerSO} table.
The same calculations are done
for the rest of $SO_{n+1}$ until all the N-dimensioned set of SOs is covered obtaining the complete
table \textit{PowerSO} shown in Figure~\ref{fig:DynSO}. 

Finally, the functions \textit{getSOMinPowerAfterAlloaction()} and \textit{getTentativeSOPlacement()} provide the minimum global power ($GlobalPower_{min}$) and the corresponding SO for which it has been obtained $SO_{Pmin}$.
So, the $SO_{Pmin}$ is used for allocating the VMs in the current time slot using Algorithm~\ref{alg:SOAlgorithm}.
\begin{algorithm} [htb]
\caption{Dynamic SO Placement Policy}
\label{alg:DynSOAlgorithm}
\begin{algorithmic}[1]
\Require {\normalsize hostList}, {\normalsize vmList}, {\normalsize SOList}
\Ensure {\normalsize $SO_{Pmin}$}
\State {\normalsize vmList}.\texttt{sortDecreasingUtilization()}
\ForAll {{\normalsize SO} \textit{in} {\normalsize SOList}}
\ForAll {{\normalsize vm} \textit{in} {\normalsize vmList}}
        \State {{\normalsize bestPlacement} $\gets$ MAX}
        \State {{\normalsize bestHost} $\gets$ NULL}
        \ForAll {{\normalsize host} \textit{in} {\normalsize hostList}}
                \If {{\normalsize host} \textit{has enough resources for} {\normalsize vm}}
                        \State {{\normalsize placement} $\gets$ {\normalsize SOvalue()}}
                        \If {{\normalsize placement} $<$ {\normalsize bestPlacement}}
                                \State {{\normalsize bestHost} $\gets$ {\normalsize host}}
                                \State {{\normalsize bestPlacement} $\gets$ {\normalsize placement}}
                        \EndIf
                \EndIf
        \EndFor
        \If {{\normalsize bestHost} $\neq$ NULL}
                \State {{\normalsize tentativeSOPlacement}.\texttt{add}({\normalsize vm}, {\normalsize bestHost})}
        \EndIf
       \EndFor
       \State {{\normalsize GlobalPowerSO} $\gets$ {\normalsize \texttt{getSOGlobalPower}(tentativeSOPlacement)}}
        \State {{\normalsize PowerSO}.\texttt{add}(SO,GlobalPowerSO)}
\EndFor
\State {{\normalsize $GlobalPower_{min}$} $\gets$ {\normalsize PowerSO}.\texttt{getSOMinPowerAfterAlloaction()}}
\State {{\normalsize $SO_{Pmin}$} $\gets$ {\normalsize PowerSO}.\texttt{getTentativeSOPlacement}($GlobalPower_{min}$)}
\State \textbf{Return:} {\normalsize $SO_{Pmin}$}
\end{algorithmic}
\end{algorithm}

\section{Modeling Metaheuristic-based SO Allocation Objectives}
\label{sec:models}
In this section we aim to obtain an expression that defines the energy
behavior of our SA allocation algorithm using local optimizations.
For this purpose, we model the global energy consumption at each
optimization slot for the metaheuristic ($Energy_{SA}$) using the
parameters of the different SO described in
Section~\ref{sec:algorithmDescription}.  The conducted experiments
have the same configuration as the ones described in
Subsection~\ref{sec:expConfig}.
First, we run the workload using the SA optimization algorithm for
VM allocation and, after each time slot, we collect the global energy
consumption $E_{SA}$.  Under the same context, we run the workload
using each $SO_{n}$ algorithm and, after each optimization slot we
monitor: i) the consolidation value normalized in the range [1,2]
($|SO_{n}|^{2}_{1}$), and ii) the total energy consumption of the
infrastructure after the consolidation process is completed
($E_{SO_n}$).  Considering the global energy of the entire data center
helps us to incorporate in the optimization the knowledge not only
from the IT and cooling contributions but also from the contributions
of the VM migrations that are needed after the allocation to avoid
underloaded situations.
In this work we use a SA configured as in Subsection~\ref{SACONFIG}.
For SA samples, we separate the entire monitored data into a training
and a testing data set.  The data set used for this modeling process
consists of the samples collected during the simulation of only the
first 24 hours of the Workload 1.  Also, we only use those samples in
which the SA outperforms the SO policies in terms of energy, as the SA
not always perform better than the SO strategies because it could
provide worse final solutions to get out from local minima.  We train
the models, using the classic regression \textit{lsqcurvefit} from MATLAB,
inferring the expressions shown in Equation~\ref{eq:SOSA}, where
$|SO_3|^{2}_{1}$ and $|SO_6|^{2}_{1}$ are the normalized consolidation
values obtained for local optimizations $SO_3$ and $SO_6$
respectively.
\begin{eqnarray}
\label{eq:SOSA}
SO_{SA} & = & E_{SA} = 0.1603 \cdot |SO_3|^{2}_{1} \cdot E_{SO_3} + 0.7724 \cdot |SO_6|^{2}_{1} \cdot E_{SO_6} + 0.0102
\end{eqnarray}

\begin{figure} [htb]
\centering
\includegraphics[width=0.7\columnwidth]{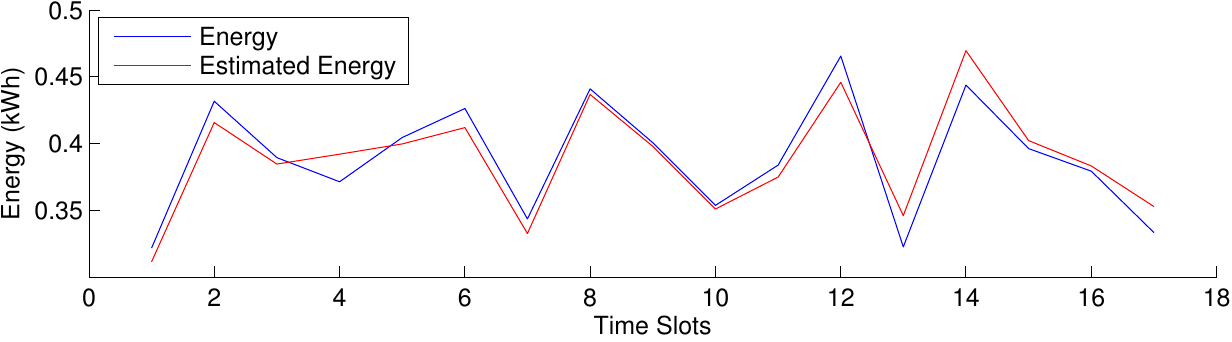}
\setlength{\belowcaptionskip}{-20pt}
\caption{Modeling fitting for $SO_{SA}$ using Simulated Annealing samples.}
\label{fig:SO9Training}
\end{figure}

\begin{figure} [htb]
\centering
\includegraphics[width=0.7\columnwidth]{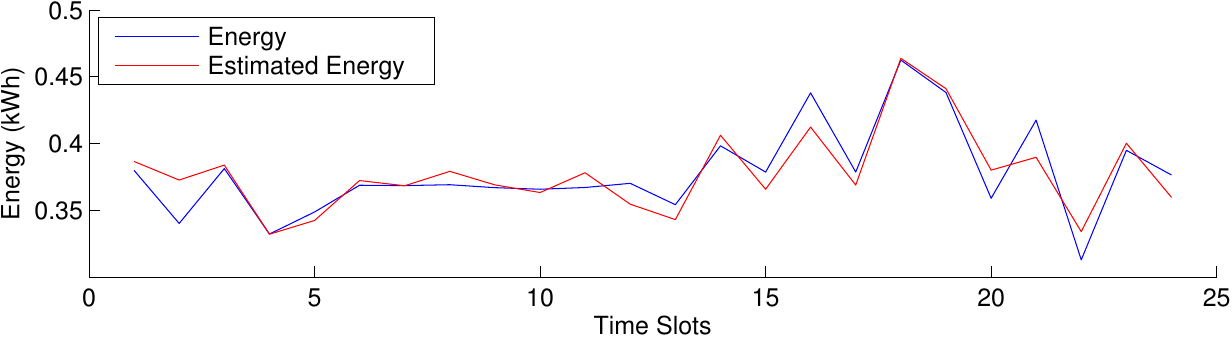}
\setlength{\belowcaptionskip}{-20pt}
\caption{Testing modeling for $SO_{SA}$  using Simulated Annealing samples.}
\label{fig:SO9Validation}
\end{figure}

The fitting is shown in Figures~\ref{fig:SO9Training}
and~\ref{fig:SO9Validation} for training and testing respectively.
For the SA energy, $E_{SA}$, we obtain an average error percentage of
3.05\% and 2.87\% for training and testing.
Finally, we use this expression to calculate
the consolidation value used in $SO_{SA}$.

\begin{figure} [htb]
\centering
\includegraphics[width=0.50\textwidth]{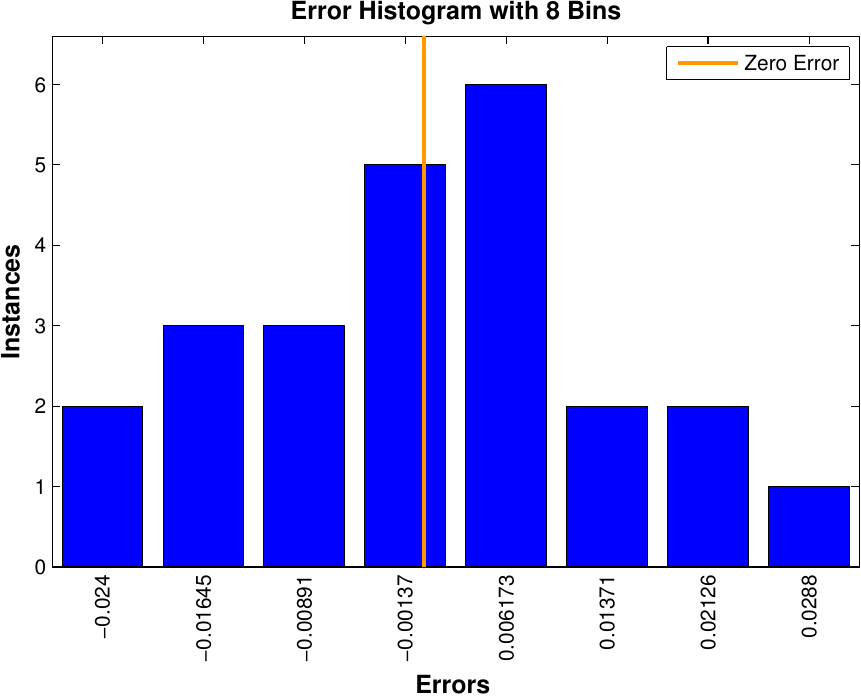}
\setlength{\belowcaptionskip}{-20pt}
\caption{Power error distribution for our $SO_{SA}$ model.}
\label{fig:ploterrhistSOSA}
\end{figure}

For our model we obtain a mean error between the estimated energy and
the real trace of $8.84 \cdot 10^{-7}$~kWh and a standard deviation of
0.0144~kWh. Figure~\ref{fig:ploterrhistSOSA} shows the power error
distribution for this model, where it can be seen that the error in
terms of power of about the 68\% of the samples may range from -0.0144
to 0.0144 kWh.

\section{Cooling strategy based on VM allocation}
\label{sec:cooling}
The power needed to cool down the servers, thus maintaining a safe temperature,
is one of the major contributors to the overall data center budget.  Many of
the reliability issues and system failures in a data center are given by the
adverse effects due to hot spots that may also cause an irreversible damage in
the IT infrastructure.  However, controlling the set point temperature of the
data room is still to be clearly defined and represents a key challenge from
the energy perspective.  This value is often chosen for the worst case scenario
(all devices running consuming maximum power), and based on conservative
suggestions provided by the manufacturers of the equipment, resulting in
overcooled facilities.  In this section we present a novel cooling strategy
based on the temperature of the system's devices due to VMs' allocation that
can be seen in Algorithm~\ref{alg:CoolingAlgorithm}.
Our cooling strategy, called \textit{VarInlet}, aims to find the highest cooling set point  of the CRAC units that ensures
a safe operation for the whole data center infrastructure under variable workload conditions.  Inside the physical
machine, the CPU is the component that presents the highest temperatures and this parameter depends
on both the inlet temperature and the CPU utilization (\textit{utilization}).  So, the CPU
temperature will limit the highest value for the inlet temperature of the host
(\textit{maxInletTemperature}) in order to operate in a safe range (lower than
\textit{maximumSafeCPUTemperature}) avoiding thermal issues.  Depending on the VMs'
distribution and the server location, we define a maximum cooling set point
(\textit{maxCoolingSetPoint}) for each host that ensures that its maximum inlet
temperature is not exceeded so the CPU temperature is safe.  Finally the cooling set point is set to the
lowest value within the \textit{maxCoolingSetPoint} for all the servers, thus
guaranteeing that the infrastructure operates below the \textit{maximumSafeCPUTemperature} in each optimization slot.

\renewcommand{\algorithmicrequire}{\textbf{Input:}}
\renewcommand{\algorithmicensure}{\textbf{Output:}}
\renewcommand{\algorithmicforall}{\textbf{foreach}}
\begin{algorithm} [htb]
\caption{Cooling strategy}
\label{alg:CoolingAlgorithm}
\begin{algorithmic}[1]
\Require {\normalsize hostList} {\normalsize maximumSafeCPUTemperature}
\Ensure {\normalsize cooling} 
       \ForAll {{\normalsize host} \textit{in} {\normalsize hostList}} 
               \State {\normalsize utilization} $\gets$ {\normalsize host}.\texttt{getUtilization()} 
                \State {\normalsize maxInletTemperature} $\gets$ {{\normalsize host}.\texttt{getMaxInletTemperature}({\normalsize utilization, maximumSafeCPUTemperature})}     
                \State {\normalsize maxCoolingSetPoint} $\gets$ {{\normalsize host}.\texttt{getMaxCoolingSetPoint}({\normalsize maxInletTemperature})}     
               \State {{\normalsize hostCooling}.\texttt{add}({\normalsize maxCoolingSetPoint})}
       \EndFor
        \State {\normalsize globalMaxCoolingSetPoint} $\gets$ {\normalsize hostCooling}.\texttt{getMin()}     
        \State {{\normalsize cooling}.\texttt{setCoolingSetPoint}({\normalsize maxCoolingSetPoint})}        
\State \textbf{Return:} {\normalsize cooling}
\end{algorithmic}
\end{algorithm}

\section{Performance Evaluation}
\label{sec:resultsTemp}

In this section, we present the impact of our proposed optimization strategies
in the energy consumption of the data center, including both IT and cooling
contributions.  As it is difficult to replicate large-scale experiments in a
real data center infrastructure, thus maintaining experimental system
conditions, we have chosen the CloudSim 2.0
toolkit~\cite{Calheiros:2011:CTM:1951445.1951450} to simulate a IaaS
(Infrastructure as a Service) Cloud computing environment.  This simulator has
been successfully used for VM allocation in Cloud computing at different
levels, taking into account multiple objectives and constraints, and using both
heuristic and metaheuristic
approaches~\cite{heilig2016cloud},~\cite{heilig2017location}.  For this work,
we have provided DVFS (Dynamic Voltage and Frequency Scaling) management and
thermal-awareness to the CloudSim simulator.  Moreover, the temperature of
servers' inlet, memory devices and CPUs vary depending on the workload
distribution and on the resource demand. We incorporate these dependence by
including different thermal models. Also our frequency and thermal-aware server
power model has been included.  Finally, in order to obtain temperature and
power performance, we have also incorporated memory and disk usage management.
Our simulations run on a 64-bit Ubuntu 14.04.5 LTS operating system running on
an Intel Core i7-4770 CPU @3.40GHz ASUS Workstation with four cores and 8 GB of
RAM. Experiments are configured according to the following considerations.

\subsection{Experimental Setup}
\label{sec:expConfig}
\begin{figure} [htb]
\centering
\includegraphics[width=0.7\columnwidth]{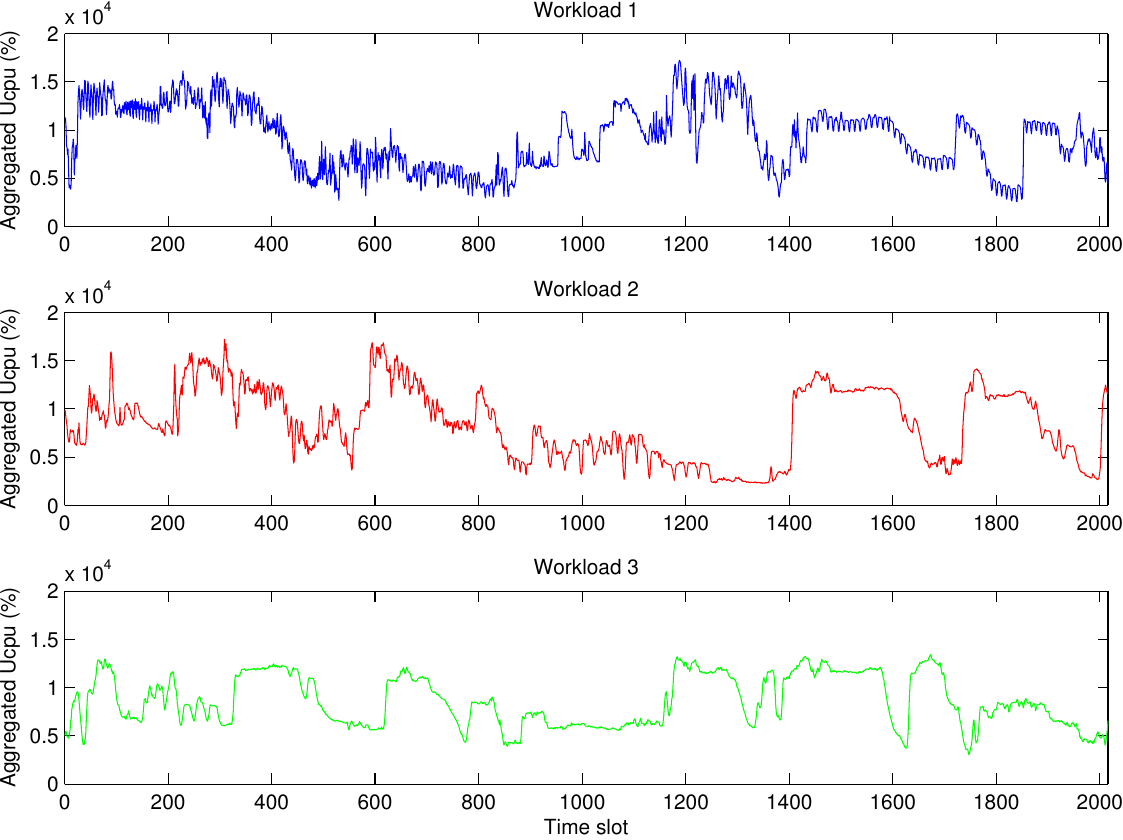}
\setlength{\belowcaptionskip}{-20pt}
\caption{One-week workloads with different CPU utilization variability.}
\label{fig:workloadProfiles}
\end{figure}

We conduct our experiments using real data from the Bitbrains service
provider.  This workload has the typical characteristics of Cloud
computing environments in terms of variability and
scalability~\cite{DBLP:conf/ccgrid/ShenBI15}.  Our data set contains
performance metrics of 1,127 VMs from a distributed data center.  It
includes resource provisioning and resource demand of CPU, RAM and
disk as well as the number of cores of each VM with a monitoring
interval of 300 seconds. These parameters define the heterogeneous VM
instances available for all the simulations.  We split the data set
into three workloads that provide three scenarios with different CPU
variability.  Each scenario represents one week of real traces from
the Bitbrains Cloud data center.  As can be seen in
Figure~\ref{fig:workloadProfiles}, Workloads 1 to 3 present decreasing
aggregated CPU utilization variability of 568.507\%, 284.626\% and
143.603\% respectively.
The simulation consists of a data center of 1200 hosts modeled as a
Fujitsu RX300 S6 server based on an Intel Xeon E5620 Quad Core
processor @2.4GHz, RAM memory of 16GB and storage of 1GB, running a
64bit CentOS 6.4 OS virtualized by the QEMU-KVM hypervisor.
During the simulations, the number of servers will be significantly
reduced as oversubscription is enabled. The proposed Fujitsu RX300 S6
server is based on an Intel Xeon E5620 processor operating at
$f_{m2}=1.73$ GHz, $f_{m3}=1.86$ GHz, $f_{m4}=2.13$ GHz, $f_{m5}=2.26$
GHz, $f_{m6}=2.39$ GHz and $f_{m7}=2.40$ GHz. For optimization
purposes, we have simulated all our algorithms under the frequency
constraints of our ad-hoc DVFS-performance aware governor proposed in
our previous work~\cite{arroba2017dynamic}, as it has been demonstrated
to give further energy improvements without affecting SLA.  Moreover,
maximum CPU temperature is constrained to take values that are lower
or equal to 65$\degree$ C for reliability purposes. Also server's
inlet is limited to a 30$\degree$ C upper bound, to avoid fan
failures.

\subsubsection{Power and Thermal Models}
\label{powAndTherModels}
In this subsection, we present the different models referred in our
previous work, and also two novel ones derived for this research. To
estimate the energy consumed by the IT infrastructure, we use our DVFS
and thermal aware server power model defined in our previous
research~\cite{Arroba:EProcedia:2014}.
\begin{eqnarray}
\label{eq:powerModel}
P_\mathrm{Fujitsu}(k,w) & = & 3.32 \cdot V^2_\mathrm{DD}(k) \cdot f_{op}(k) \cdot u_\mathrm{CPU}(k,w) + 1.63 \cdot 10^{-3} \cdot T^2_\mathrm{MEM}
+ 4.88 \cdot 10^{-11} \cdot FS^3 \\
\label{en}
E_\mathrm{Fujitsu}(k,w) & = & P_\mathrm{Fujitsu}(k,w) \cdot t \\
\label{freqs}
f_{op}(k) & \in & \{1.73, 1.86, 2.13, 2.26, 2.39, 2.40\}\\
\label{bootingenergy}
E_{boot} & = & 13.514 \cdot 10^{-3} kW\cdot h
\end{eqnarray}

Equation~\ref{eq:powerModel} shows the power consumption of the
Fujitsu server, where $V_\mathrm{DD}(k)$ is the CPU supply voltage and
$f_{op}(k)$ is the working frequency in GHz of the server in a
specific \textit{k} DVFS mode. $u_\mathrm{cpu}(k,w)$ is the averaged CPU
percentage utilization running a workload \textit{w} at the \textit{k} DVFS mode.
$T_\mathrm{mem}$ defines the memory temperature in Kelvin and \textit{FS}
describes the fan speed in RPM.
This model achieves a testing error of 4.46\% when comparing power
estimation to real measurements of the actual power in Cloud
applications. The energy consumption is obtained using
Equation~\ref{en} where \textit{t} defines the time in which the energy value
is required. In our research, the time slots are defined as each time
an optimization is performed in order to consolidate a set of VMs into
a set of candidate hosts. We use a value of \textit{t} of 300 seconds to
match the workload traces provided by BitBrains. The operating
frequencies set (in GHz) is provided in~\ref{freqs}.  In this work, we
assume that servers are powered on and off when needed, so we take
into account the booting energy consumption required by a server to be
fully operative as shown in Equation~\ref{bootingenergy}.

As our power model shown in Equation~\ref{eq:powerModel} depends on
the memory temperature, for this research, we present a novel
temperature model for the memory device.
The temperature of the memory device in a server depends on several
factors both internal and external to the physical machine. The
utilization of the memory subsystems, the inlet temperature of the
host and the fan speeds are potential contributors to memory
temperature that have to be taken into account.  In order to gather
the real data during runtime, we monitor the system using different
hardware and software resources.  \textit{collectd} monitoring tool is used
to collect the values taken by the system in order to monitor
$u_{MEM}$.  Memory and CPU temperatures and fan speed are monitored
using on board sensors that are consulted via the \textit{IPMI} software
tool.  Inlet temperature is collected using external temperature
sensors. Finally, room temperature has been modified during run-time
in order to find the dependence with the inlet temperature.

In this research, a synthetic workload is used to stress specifically
the memory resources, increasing the range of possible values of the
considered variables. Therefore, our model may be adapted to estimate
different workload characteristics and profiles.  
We run a version of
$RandMem$ \footnote{http://www.roylongbottom.org.uk} (modified to
access random memory regions individually) onto 4 parallel Virtual
Machines that have been provisioned to the available computing
resources of the server. Then the samples are separated into a
training and a testing data set.
After training, we obtain the model shown in
Equation~\ref{eq:modelTmem}, where $T_{mem}$ is the memory temperature,
$U_{mem}$ the memory utilization, $T_{inlet}$ the inlet temperature of the
server, $k_1 = 0.9965$ and $k_2 = 2.6225$. Then, we evaluate the
quality of the thermal model using the testing data set in order to
verify the reliability of the estimation.
For our data fitting, we obtain an average error percentage of
0.5303\% during training and 0.5049\% for testing. These values have
been obtained using Equation~\ref{eq:avgerr}.
\begin{eqnarray}
T_{mem} & = & k_1 \cdot T_{inlet} + k_2 \cdot ln(U_{mem}^2) \label{eq:modelTmem}\\
\label{eq:avgerr}
e_{AVG} & = & \sqrt{\frac{1}{N} \cdot \sum_{n} {\Big(\frac{| T_{mem}(n) - \widehat{T_{mem}}(n) | \cdot 100} {T_{mem}(n)}\Big)}^2}, 1 \leq n \leq N 
\end{eqnarray}

\begin{figure} [htb]
\centering
\includegraphics[width=0.50\textwidth]{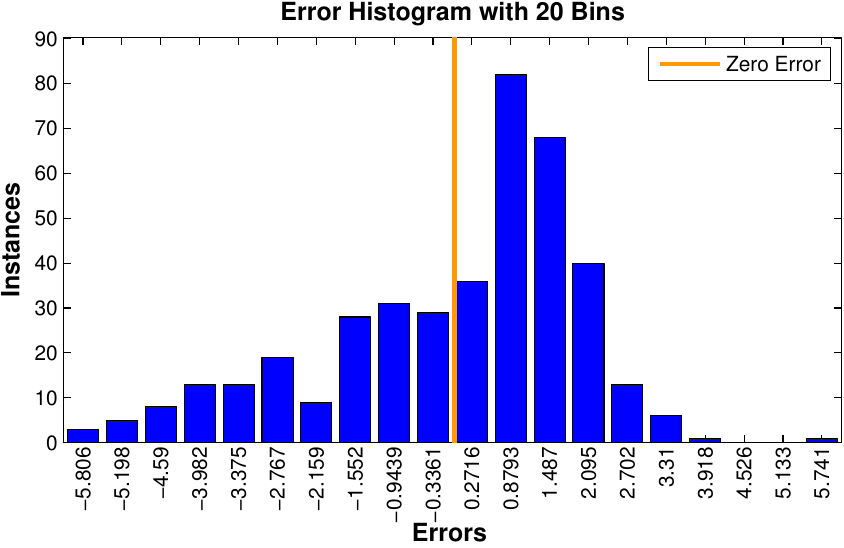}
\setlength{\belowcaptionskip}{-20pt}
\caption{Temperature error distribution for our memory model.}
\label{fig:ploterrhistTmem}
\end{figure}

\begin{figure} [htb]
\centering
\includegraphics[width=0.5\columnwidth]{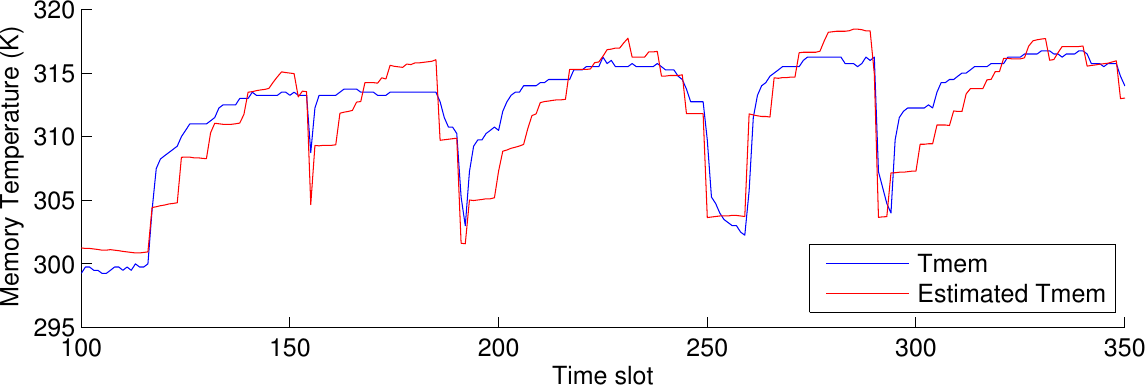}
\setlength{\belowcaptionskip}{-20pt}
\caption{Modeling fitting for the memory temperature.}
\label{fig:TmemTraining}
\end{figure}

Finally, for our thermal model we obtain a mean error between the
estimated temperature and the real measurement of $8.54 \cdot
10^{-4}$~K and a standard deviation of
2.02~K. Figure~\ref{fig:ploterrhistTmem} shows the error distribution
for this model. According to this, we can conclude that the error in
terms of temperature of about the 68\% of the samples ranges from
-2.02 to 2.02~K.  In Figure~\ref{fig:TmemTraining}, the fitting of our
thermal model is provided.

\begin{figure} [htb]
\centering
\includegraphics[width=0.50\textwidth]{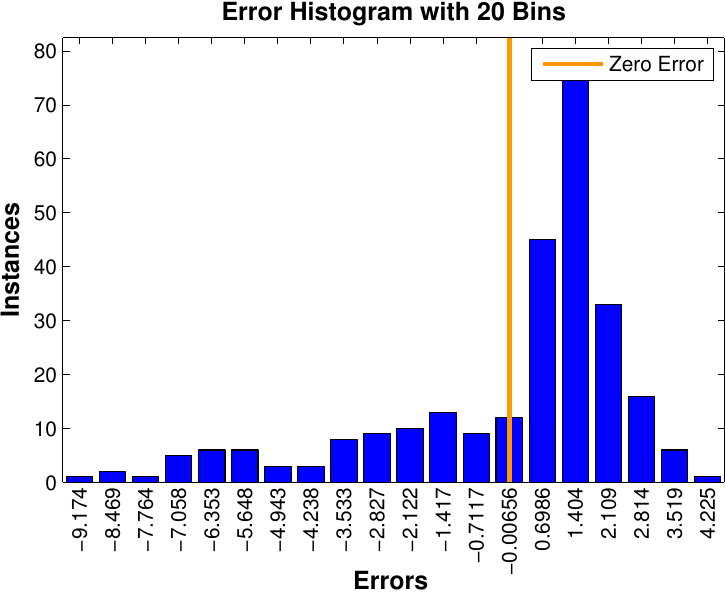}
\setlength{\belowcaptionskip}{-20pt}
\caption{Temperature error distribution for our CPU model.}
\label{fig:ploterrhistTcpu}
\end{figure}

\begin{figure} [htb]
\centering
\includegraphics[width=0.5\columnwidth]{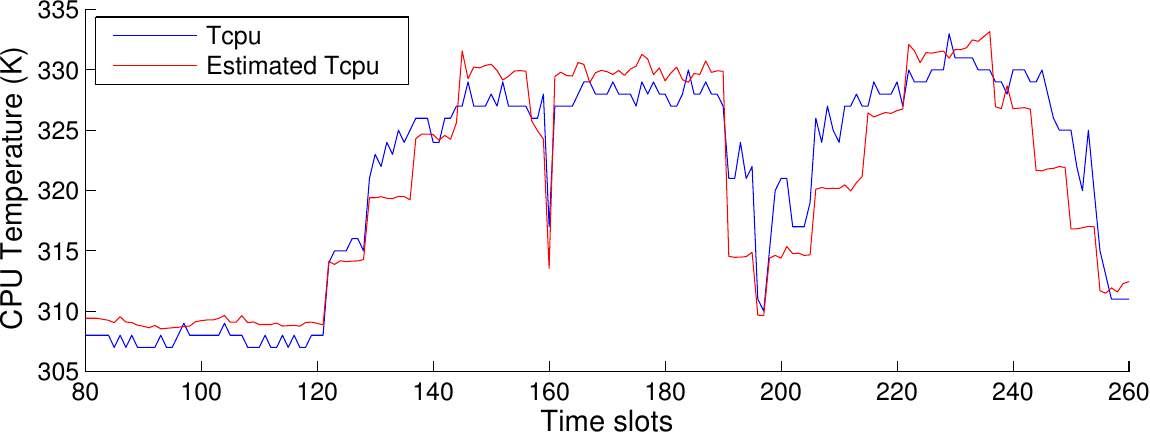}
\setlength{\belowcaptionskip}{-10pt}
\caption{Modeling fitting for the CPU temperature.}
\label{fig:TcpuTraining}
\end{figure}

On the other hand, the CPU presents the highest temperatures inside
the physical machine, so its temperature will limit the highest value
for inlet temperature in order to operate in a safe range, while
avoiding thermal issues.  Thus, we follow the same approach in order
to model the CPU temperature of the server.  This parameter depends on
both the inlet temperature and the CPU utilization ($u_{CPU}$).
After training, we obtain the model shown in
Equation~\ref{eq:modelTcpu}, where $T_{cpu}$ is the CPU temperature,
$U{cpu}$ its utilization, $k_1 = 1.052$ and $k_2 = 19.845$.  Our model
presents average error percentages of 0.64\% and 0.84\% during
training and testing respectively.
\begin{eqnarray}
T_{cpu} & = & k_1 \cdot T_{inlet} + k_2 \cdot U_{cpu} \label{eq:modelTcpu}
\end{eqnarray}
Finally, we obtain a mean error between the estimated temperature and
the real measurement of 0.0026~K and a standard deviation of
2.683~K. Figure~\ref{fig:ploterrhistTcpu} shows the error distribution
for this model, where the error in terms of temperature of about the
68\% of the samples ranges from -2.68 to 2.68~K.  In
Figure~\ref{fig:TcpuTraining}, the fitting of our thermal model is
provided.
Disk power consumption is modeled according to the work proposed by
Lewis et al.~\cite{Lewis:2008:REC:1855610.1855614}, as can be seen in
Equation~\ref{eq:pdisk}, where $Disk_{r}$ and $Disk_{w}$ are the read
and write throughputs.  We define cooling energy model, shown in
Equation~\ref{eq:coolingPower}, as in the research presented by Moore
et al.~\cite{Moore:2005:MSC:1247360.1247365}.  The COP depends on the
inlet temperature of the servers' $T_{inlet}$.
\begin{eqnarray}
\label{eq:pdisk}
P_{Disk} & = & 3.327 \cdot 10^{-7} \cdot Disk_{r} + 1.668 \cdot 10^{-7} \cdot Disk_{w} \\
\label{eq:coolingPower}
E_{Cooling} & = & E_{IT} / COP = (P_{IT} \cdot t) / COP \\
E_{IT} & = & (P_{Fujitsu} + P_{Disk}) \cdot t \\
COP & = & 0.0068 \cdot T_{inlet}^2 + 0.0008 \cdot T_{inlet} + 0.458 
\end{eqnarray}

\subsubsection{Dynamic Consolidation considerations}
In all our scenarios we allow online migration, where VMs follow a
straightforward load migration policy.  Thus, migrations have an
energy overhead because, during the migration time, two identical VMs
are running, consuming the same power in both servers.  Performance
degradation occurs when the workload demand in a host exceeds its
resource capacity.
Our dynamic consolidation strategy first chooses which VMs have to be
migrated in each server of the data center.  For this purpose, we use
the adaptive utilization threshold based on the Median Absolute
Deviation (MAD) of the CPU and the Minimum Migration Time (MMT)
algorithm provided by Beloglazov et
al.~\cite{Beloglazov:2012:OOD:2349876.2349877}. Then, the VM allocation
is performed according to the optimization algorithms provided in
Section~\ref{sec:algorithmDescription}.
In this work we allow oversubscription in all the servers so, the
total resource demand may exceed their available capacity.  If the VMs
in a host simultaneously request their maximum performance, this
situation may lead to performance degradation due to host overloading.
When overloading situations are detected using the MAD-MMT technique,
VMs are migrated to better placements according to the different
proposed algorithms.  This procedure creates a performance
degradation due to the migrations required. In this research, we
determine SLA violations ($SLA_{violation}$), as the metric provided
by CloudSim, further explained in the research proposed by Beloglazov
et al.~\cite{Beloglazov:2012:OOD:2349876.2349877}.  This metric
combines the performance degradation due to both host overloading and
migrations, and would help us to monitor the SLA provided by our
optimization system.

\subsubsection{SA Configuration}
\label{SACONFIG}
In this work we use the SA provided by the HEuRistic
Optimization (HERO) library of optimization
algorithms\footnote{github.com/jlrisco/hero}. We executed 32 simulations to
determine the maximum number of iterations, as well as the value of k. After
these tests we observed that the best solution was never improved beyond 100000
iterations, independently of the value of k. k is a weight used to compute the
probability of changing the state to a new solution. $k=0.5$ offered the best
solutions overall. The probability of changing the state when the new solution
worsens the current solution depends on $e^{-(x/k)}$ so, for $k=0.5$, it
results on $e^{-(2x)}$ meaning that the probability of changing the state is
reduced in such exponential factor. This makes sense when the initial solution
is “good” (we set the initial solution to the best found by the SO policies)
because an improvement in energy can be reached with small number of
permutations of virtual machines. Additionally, we had a hard time constraint
for performance purposes, for instance, the maximum execution time of the SA
algorithm should not exceed 300 seconds, and 100000 iterations safely entered
in that time window.  Finally, we modified the basic implementation of the HERO
library to facilitate the selection of the best-so-far solution, which is not
included by default.

\subsection{Baselines}
\label{baselines}
In this research, we compare our proposed technique against a heuristic, a
metaheuristic, and a recent state-of-the-art algorithm. As heuristic, we select
a BFD-based algorithm due to its light implementation in terms of execution
time and complexity, still being powerful enough to tackle optimization
problems like the one described; as metaheuristic, we select an optimizer based on gradient descent due to its global minimization
capabilities.
According to the state-of-the-art that we present in Section 2,
we would like to compare our proposed algorithms with a heuristic.  The PABFD
approach, proposed by Beloglazov et
al.~\cite{Beloglazov:2012:OOD:2349876.2349877}, is used as our BFD-based
heuristic baseline.  
Moreover, this BFD algorithm is single-objective and provides a local optimization, so it is fast and light to be used during run-time.
This algorithm is referred in many recent
publications~\cite{Chowdhury2015},~\cite{7776481},~\cite{Solanki},~\cite{8226661},~\cite{Mosa2016},~\cite{10.1007/978-3-319-26135-5_15},~\cite{Quang-Hung:2016:EVM:3081105.3081111},
and it is also included in the open source version of CloudSim 2.0.
The second baseline is an in-house simulated annealing
approach. This algorithm is single-objective and provides a global optimization
of the data center energy consumption for both IT and cooling contributions. We
have enhanced the SA, initializing the algorithm to the best solution found by
the SO set of algorithms. This initialization ensures that the SA finds a
feasible solution that is, at least, as good as the one provided by the best SO
in each optimization slot.
Both PABFD and SA have been explained thoroughly in
Sections~\ref{SO1} and ~\ref{SAsect}, as they have been used as objectives to
define our multi-objective policies ($MO_1$ and $MO_2$) and to model our
metaheuristic-based single-objective approach ($SO_{SA}$) respectively.
Finally, we include the SWFDVP (Second worst fit decreasing VM
placement) algorithm as baseline, which has been recently proposed by Chowdhury
et al.~\cite{Chowdhury2015}.  This single-objective BFD policy is based on the almost worst fit
decreasing technique and aims to allocate a VM in the bin that provides the
second minimum empty space. More in detail, the objective of the SWFDVP
algorithm is to maximize the power increment due to the allocation of a VM
within a set of hosts, and select the host that provides the second best
solution. This baseline helps us to compare our work with a real-time state-of-the-art research.

\subsection{Experimental results}
\begin{figure} [htb]
\centering
\includegraphics[width=0.7\columnwidth]{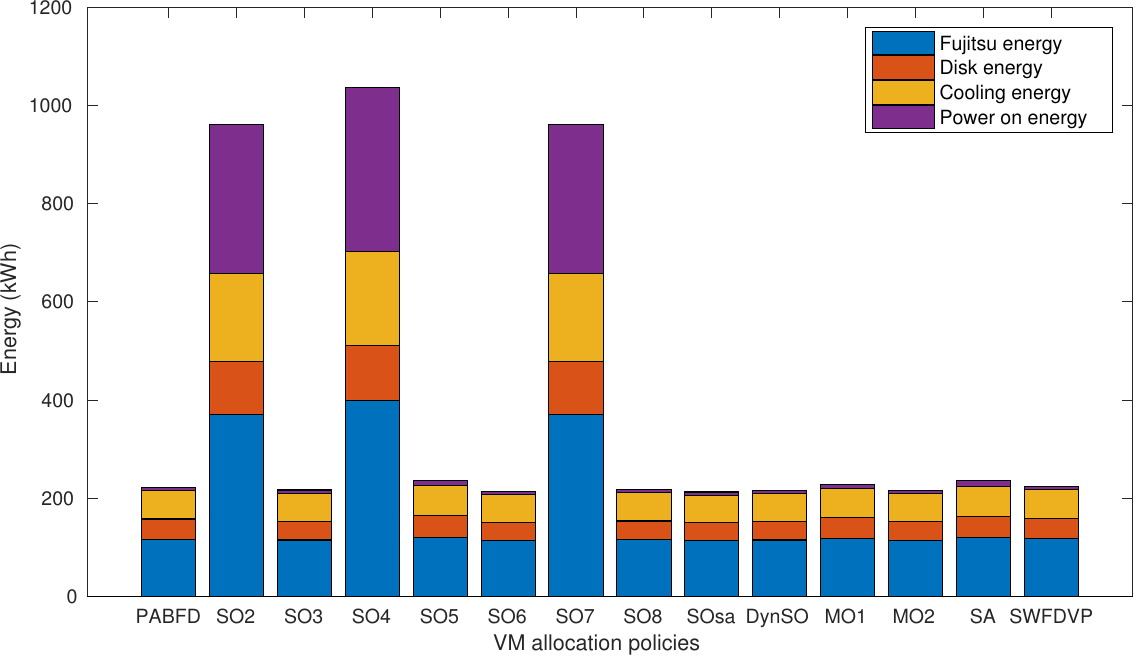}
\setlength{\belowcaptionskip}{-10pt}
\caption{Contributions to data center energy per VM allocation
  strategy for 1 day of Workload 1}
\label{fig:bar1}
\end{figure}

To obtain a preliminary evaluation of the performance of the different
VM allocation policies, we have simulated one day of our Workload 1
from Bitbrains using the proposed strategies presented in
Section~\ref{sec:algorithmDescription} for a fixed inlet temperature
of 291K.  On the one hand, the simulation time when using the SO
approaches ($SO_{1-8}$, $SO_{SA}$, and \textit{DynSO}) ranges from 8 to 12
minutes. This parameter for MO approaches is around 15 minutes, being
in the same order of magnitude. On the other hand, the
metaheuristic-based SA has a simulation time that is 60 times higher
than the complete simulation time of SO policies.  Also, for SA, there
exist some optimizations that take more time than the time fixed for
the optimization slot (300~s), making it unfeasible to use this
metaheuristic during runtime. The energy results provided by the
selected algorithms are shown in Figure~\ref{fig:bar1}.  Also,
Table~\ref{tab:291K_1day} shows the numerical results of the
additional metrics considered.

\begin{table*} [htb]
\scriptsize
\centering
\caption {Performance metrics per VM allocation strategy for 1 day of
  Workload 1}
\begin{tabular}{lrrrrrrr}
  \hline
  Algorithm & IT Energy & Cooling Energy & Power-on & Power-on Energy    & Migrations & Average SLA     & Final Energy \\
  	    & (kWh)	& (kWh)	         & events	  & (kWh) 	       & events	    & $\cdot 10^{-4}$ (\%) & (kWh) \\
  \hline
PABFD $(SO_1)$       & 157.62 &	58.91	& 309	& 5.80		& 42545		& 18.43	& 222.32 \\ 
$SO_2$       & 478.54 &	178.85	& 16194	& 303.74	& 116044	& 11.66	& 961.13 \\
$SO_3$       & 153.15 &	57.24	& 335	& 6.28		& 36711		& 18.28	& 216.67 \\
$SO_4$       & 511.60 & 191.21	& 17792	& 333.71	& 125807	& 11.89	& 1036.52 \\
$SO_5$       & 165.12 &	61.71	& 524	& 9.83		& 45044		& 18.36	& 236.66 \\
$SO_6$       & 150.43 &	56.22	& 336	& 6.30		& 35619		& 19.29	& 212.95 \\
$SO_7$       & 478.54 &	178.85	& 16194	& 303.74	& 116044	& 11.66	& 961.13 \\
$SO_8$       & 153.54 &	57.38	& 334	& 6.26		& 37376		& 18.28	& 217.19 \\
$SO_{SA}$    & 150.20 & 56.14	& 333	& 6.25		& 36983		& 19.50	& 212.58 \\
$DynSO$	     & 152.15 & 56.86	& 326	& 6.11		& 36126		& 18.71 & 215.13 \\
$MO_1$       & 160.41 &	59.95	& 367	& 6.88		& 40342		& 18.15	& 227.25 \\
$MO_2$       & 152.86 &	57.13	& 331	& 6.21		& 37690		& 18.62	& 216.20 \\
$SA$         & 162.95 &	60.90	& 662	& 12.42		& 50540		& 18.34	& 236.27 \\
SWFDVP     & 157.95 & 59.03	& 369 	& 6.92 		& 39077 	& 18.51 & 223.90 \\
  \hline
  \end{tabular}
  \label{tab:291K_1day}
\end{table*}

For $SO_2$, $SO_4$ and $SO_7$, the power or temperature of each server
is minimized locally resulting in a higher energy consumption due to
an increase in the number of active hosts.  These algorithms spread
the workload as much as possible through the candidate host set as
they intend to reduce only the dynamic contribution, which depends on
the workload requirements. So, the lower the servers' load, the better
for reducing dynamic power consumption locally, thus increasing the
global IT contribution as these policies are not aware of their impact
on the rest of the infrastructure.  Then, after allocating those VMs
incoming from overloaded servers, in the next iteration, the algorithm
constrains the active server set by migrating VMs from underutilized
hosts if possible.  So, these algorithms present a higher number of
migrations.

Moreover, the energy consumption of both $SO_5$ and SA policies is
above the average due to a higher number of VM migrations and power on
events, performed to find the best data center configuration in each
optimization slot. This is due to their trend towards allocating part
of the workload in underutilized servers.  In the case of $SO_5$, this
situation occurs because, when servers present an utilization below
72\% (equivalent utilization for 1.73GHz that is the lowest available
frequency), their frequency increment is zero. On the other hand, for
the SA algorithm, the optimization value is highly penalized when the
solutions include overloaded servers. So, underutilized servers are
preferred when the algorithm does not find a minimum.
Our results show that $SO_3$ and $SO_6$ are the best simple-SO
optimization policies. This outcome is consistent with the high idle
consumption of the Fujitsu's server architecture, so reducing the
active servers' set by increasing CPU utilization is a major target to
improve energy efficiency.
The multi-objective strategies that we present in this research also
outperform the baseline, where $MO_2$ is more competitive in terms of
energy.

The $SO_{SA}$ strategy shows the lowest total consumption value,
providing energy savings of 4.38\%, 10.02\% and 5.06\% on the global power
budget when compared with our baselines PABFD, SA and SWFDVP
respectively. This is translated into a reduction of 9.74~kWh,
23.69~kWh and 11.32~kWh as this novel technique also incorporates global information
regarding the effect of allocation on future VM migrations.  This
local approach takes advantage of global knowledge from joint IT and cooling
viewpoint thus outperforming other strategies for highly variable
workloads.
The \textit{DynSO} policy, which dynamically sets the best SO during runtime,
also reduces power significantly, but does not achieve the best result
in terms of energy consumption.  This is because, local policies may
offer better consolidation values for the server scope but do not
consider the impact on the data center as a whole once the allocation
is performed (e.g. future migrations).  On the other hand, the
$SO_{SA}$ provides the best results in this scenario. This novel
approach may not provide the best consolidation value during local
calculations, but is the one that best describes the energy behavior
of the entire infrastructure, considering also future migrations.
Moreover, the SLA is maintained for all the tests, where a higher
increment of $1.07 \cdot 10^{-4} \%$ is detected. \\

After this proof of concept, we use all the different VM allocation
policies provided in Section~\ref{sec:algorithmDescription} to
optimize the power consumption of our proposed data center
infrastructure under several workload and cooling conditions.  In this
work we propose the optimization of 9 scenarios, running the three
different 7 days-workload profiles shown in Figure~\ref{fig:workloadProfiles}
and applying three different cooling strategies.  The cooling policies
used in this research are: i) 291K fixed setpoint temperature (found
in traditional data centers), ii) 297K fixed setpoint temperature
(found in new data center infrastructures) and iii) our \textit{VarInlet}
cooling strategy provided in Section~\ref{sec:cooling}.
According to our problem, the number of VMs to allocate is
\textit{numVMs=1127} and the number of hosts \textit{numHosts=1200}.  BFD-based
SO and MO policies ($SO_1$ to $SO_8$, \textit{DynSO}, $SO_{SA}$, $MO_1$ and
$MO_2$) are deterministic approaches with a space of solutions of
\textit{numVMs*numHosts}, so all the executions provide the same result. For this reason
we run each of them once. On the other hand, the SA is a metaheuristic that
provides different solutions that are highly dependent with the initial random
solution. In our proposed SA, the initial solution is deterministic, as it is
always initialized to the best solution provided by the set of SO policies,
thus limiting its randomness. However, the computing time of our SA scenario is
around 3-4 days due to the increased space of solutions, $numHosts^{numVMs}$ in this
case. The high run time makes it difficult to run the SA scenario for a high
number of executions, so the set of executions is not extensive enough to
provide meaningful statistics. For these reasons, we decided to run each SA
scenario only once.

First, we optimize Workload 1, which is the workload that presents the
higher instantaneous variability, for fixed cooling inlets of 291~K
and 297~K, and for our variable inlet cooling strategy
(\textit{VarInlet}). We obtain the results provided in Table~\ref{tab:WL1} in
terms of final energy consumption, average SLA and number of
migrations.

\begin{table} [phtb]
\begin{center}
\caption{Energy, SLA and Migration metrics per inlet temperature and
  allocation policy for workload 1.}
\label{tab:WL1}
\begin{tabular}{lrrrrrrrrr}
\hline
\multirow{2}{*}{Policy} & \multicolumn{3}{c}{Energy (kWh)} & \multicolumn{3}{c}{Average SLA ($\cdot 10^{-4}$ \%)} & \multicolumn{3}{c}{Migrations ($\cdot 10^{3}$)} \\
& 291K & 297K & VarInlet & 291K & 297K & VarInlet & 291K & 297K & VarInlet \\
\hline
PABFD & 1178.41	& 1089.27	& 1034.20 & 20.76 & 20.98 &	21.04 &	200.8 &	202.2 &	203.2 \\ 
$SO_2$ & 5256.76	& 4718.89	& 4594.32 & 11.97 & 11.94 &	11.88 &	647.4 &	647.7 &	654.1 \\
$SO_3$ & 1159.07	& 1059.19	& 1018.19 & 20.60 & 20.60 &	20.60 &	190.2 &	190.2 &	190.2 \\
$SO_4$ & 5725.25	& 5207.93	& 4990.10 & 11.87 & 11.87 &	11.88 &	766.6 &	766.6 &	772.2 \\
$SO_5$ & 1247.97	& 1139.50	& 1094.45 & 20.15 & 20.15 &	20.15 &	217.5 &	217.5 &	217.5 \\
$SO_6$ & 1157.61	& 1057.85	& 1016.91 & 20.90 & 20.90 &	20.90 &	189.1 &	189.1 &	189.1 \\
$SO_7$ & 5256.76	& 4718.89	& 4594.32 & 11.97 & 11.94 &	11.88 &	647.4 &	647.7 &	654.1 \\
$SO_8$ & 1161.44	& 1061.33	& 1020.22 & 20.31 & 20.31 &	20.31 &	192.5 &	192.5 &	192.5 \\
$SO_{SA}$ & 1152.13	& 1052.93	& 1012.30 & 20.84 & 20.84 &	20.84 &	187.9 &	187.9 &	187.9 \\
$DynSO$ & 1164.30	& 1055.83	& 1020.59 & 20.56 & 20.61 &	20.56 &	190.8 &	189.0 &	192.9 \\
$MO_1$ & 1199.55	& 1090.12	& 1047.56 & 20.01 & 20.31 &	19.94 &	198.0 &	199.0 &	201.0 \\
$MO_2$ & 1159.39	& 1059.45	& 1018.42 & 20.62 & 20.62 &	20.62 &	191.7 &	191.7 &	191.7 \\
$SA$   & 1293.51	& 1178.35	& 1130.98 & 19.94 & 19.54 &	19.80 & 244.9 & 241.3 & 244.6 \\
SWFDVP & 1193.87	& 1092.76	& 1051.76 & 20.60 & 21.11 &	20.44 & 188.4 & 190.9 & 185.3 \\
\hline
\end{tabular}
\end{center}
\end{table}

\begin{figure} [htb]
\centering
\includegraphics[width=0.8\columnwidth]{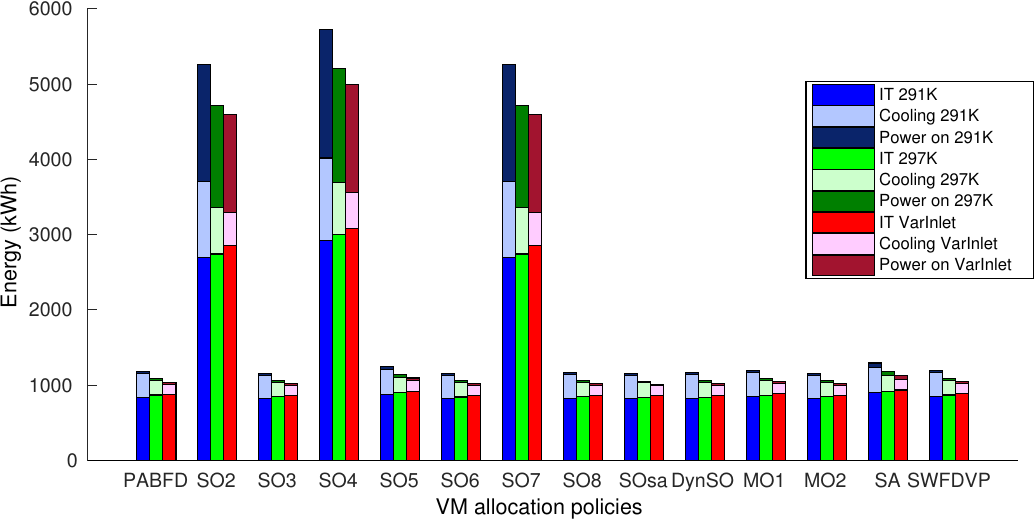}
\setlength{\belowcaptionskip}{-20pt}
\caption{Contributions to data center energy per VM allocation
  strategy for Workload 1}
\label{fig:stackedWL1}
\end{figure}

Figure~\ref{fig:stackedWL1} shows the different contributions to final
energy per VM allocation policy for the different cooling strategies.
As inlet temperature rises, the IT power consumption is increased due
to power leakage (see IT 291K, IT 297K and IT \textit{VarInlet} at the bottom
of each stacked column).  However, cooling power is reduced with
increasing temperatures due to a higher cooling efficiency (shown in
Cooling 291K, Cooling 297K and Cooling \textit{VarInlet} in the middle of each
stacked column). The savings performed by higher cooling set points
outperform the IT power increments, thus resulting in more efficient
scenarios for all the proposed allocation strategies. The energy
needed to power on the servers when it is required by the allocation
policies is shown as \textit{Power on 291K}, \textit{Power on 297K} and \textit{Power on
VarInlet} respectively for the three cooling strategies.  For the
three scenarios running Workload 1, only by applying our \textit{VarInlet}
cooling strategy provides additional energy savings of 3.78\% and
12.38\% in average when compared with fixed cooling at 297~K and 291~K
respectively for all the allocation policies.

\begin{figure} [htb]
\centering
\includegraphics[width=0.6\columnwidth]{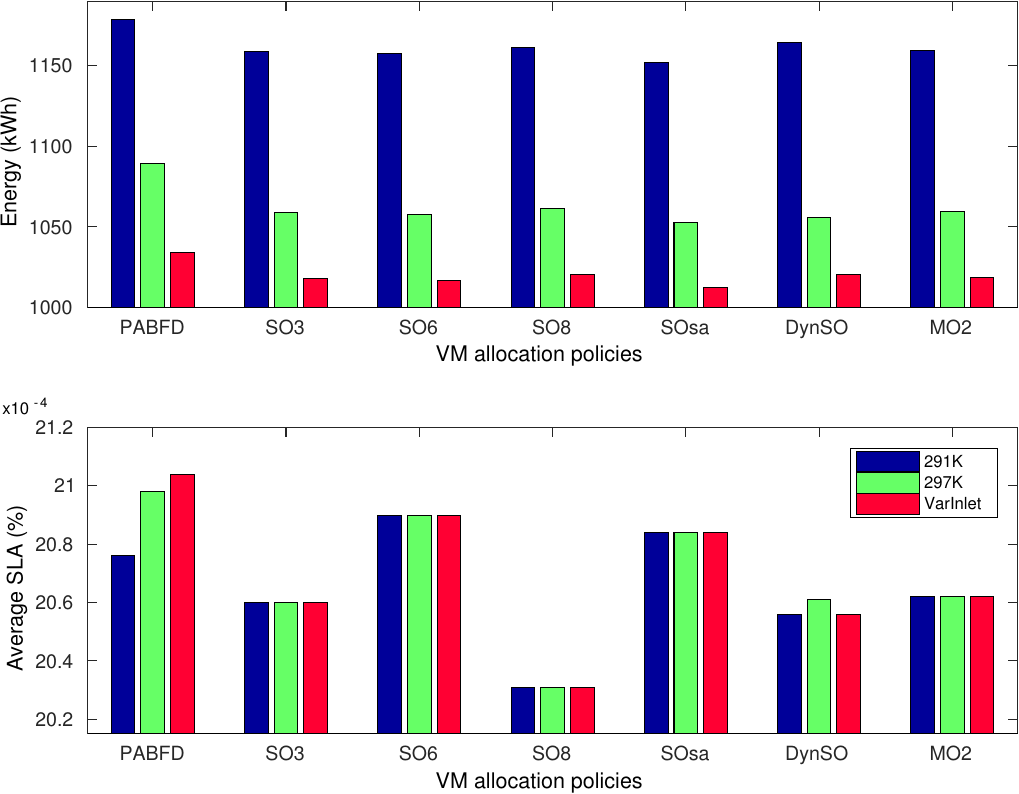}
\setlength{\belowcaptionskip}{-20pt}
\caption{Data center energy and SLA per VM allocation strategy for Workload 1}
\label{fig:nonstackedWL1}
\end{figure}

Figure~\ref{fig:nonstackedWL1} shows the total energy and average SLA
percentage comparison for those strategies that outperform the PABFD
baseline, which is the baseline that performs better in this scenario in terms of energy.  $SO_{SA}$, $SO_6$, $SO_3$, $MO_2$ and \textit{DynSO} allocation
policies offer better results, in terms of energy savings, when
compared to our global baseline SA and our local baselines
PABFD and SWFDVP. These policies, when combined with \textit{VarInlet} strategy,
provide savings, of 13.67\% and 21.35\% in average with respect to
SA at 297~K and 291~K respectively. Maximum savings are found of up
to 14.09\% and 21.74\% respectively for our \textit{VarInlet}-$SO_{SA}$
combined strategy.  When compared with the local baseline PABFD,
these allocation policies provide average savings of 6.61\% and
13.67\% at 297~K and 291~K respectively, and maximum savings of up to
7.07\% and 14.10\% for $SO_{SA}$.
Finally, comparing the results with the SWFDVP baseline, average savings of 6.91\% and 14.79~\% and maximum savings of 7.33~\% and 15.21\% (for $SO_{SA}$) are found.

In our three following scenarios we optimize Workload 2, which
presents medium instantaneous variability.  We obtain the energy
consumption, average SLA and migration results provided in
Table~\ref{tab:WL2}.
\begin{table} [htb]
\begin{center}
\caption{Energy, SLA and Migration metrics per inlet temperature and allocation policy for workload 2.}
\label{tab:WL2}
\begin{tabular}{lrrrrrrrrr}
\hline
\multirow{2}{*}{Policy} & \multicolumn{3}{c}{Energy (kWh)} & \multicolumn{3}{c}{Average SLA ($\cdot 10^{-4}$ \%)} & \multicolumn{3}{c}{Migrations ($\cdot 10^{3}$)} \\
& 291K & 297K & VarInlet & 291K & 297K & VarInlet & 291K & 297K & VarInlet \\
\hline
PABFD & 1033.75	& 949.09	& 912.18	& 18.84	& 18.67	& 18.58	& 119.3	& 126.0	 & 123.0 \\ 
$SO_2$ & 3712.18	& 3422.65	& 3278.08	& 11.15	& 11.11	& 11.21	& 564.2	& 579.4	 & 570.9 \\
$SO_3$ & 1022.49	& 935.02	& 899.36	& 18.43	& 18.43	& 18.43	& 121.7	& 121.7	 & 121.7 \\
$SO_4$ & 4562.67	& 4142.05	& 3978.78	& 11.00	& 11.00	& 11.01	& 751.5	& 751.5	 & 761.0 \\
$SO_5$ & 1081.89	& 988.79	& 950.59	& 18.11	& 18.11	& 18.11	& 136.0	& 136.0	 & 136.0 \\
$SO_6$ & 1015.27	& 928.55	& 893.27	& 18.51	& 18.51	& 18.51	& 120.8	& 120.8	 & 120.8 \\
$SO_7$ & 3712.18	& 3422.65	& 3278.08	& 11.15	& 11.11	& 11.21	& 564.2	& 579.4	 & 570.9 \\
$SO_8$ & 1024.57	& 936.92	& 901.17	& 18.46	& 18.46	& 18.46	& 120.3	& 120.3	 & 120.3 \\
$SO_{SA}$ & 1016.71	& 929.82	& 894.48	& 18.54	& 18.54	& 18.54	& 119.3	& 119.3	 & 119.3 \\
$DynSO$ & 1027.06	& 939.13	& 903.27	& 18.24	& 18.24	& 18.24	& 118.9	& 118.9	 & 118.9 \\
$MO_1$ & 1032.37	& 945.82	& 913.50	& 18.21	& 18.16	& 18.21	& 125.0	& 121.8	 & 122.6 \\
$MO_2$ & 1024.80	& 937.07	& 901.30	& 18.28	& 18.28	& 18.28	& 121.5	& 121.5	 & 121.5 \\
$SA$   & 1122.67	& 1026.40	& 985.68	& 18.34 & 18.12 & 18.06	& 160.1 & 163.3	 & 161.6 \\
SWFDVP & 1055.37	& 967.99	& 931.64	& 18.36 & 18.36 & 18.85 & 118.3 & 120.1  & 120.8 \\
\hline
\end{tabular}
\end{center}
\end{table}
In Figure 18
, the same trend towards power and
temperature is shown as in Workload 1 scenarios.  Savings obtained by
increasing cooling set points also outperform IT power increments,
thus resulting in more efficient scenarios for all the proposed
allocation strategies.  For this scenario, only our \textit{VarInlet} cooling
strategy provides additional energy savings of 3.88\% and 12.00\% in
average, when compared with fixed cooling at 297~K and 291~K
respectively, for all the allocation policies.

\begin{figure} [phtb]
\centering
\includegraphics[width=0.8\columnwidth]{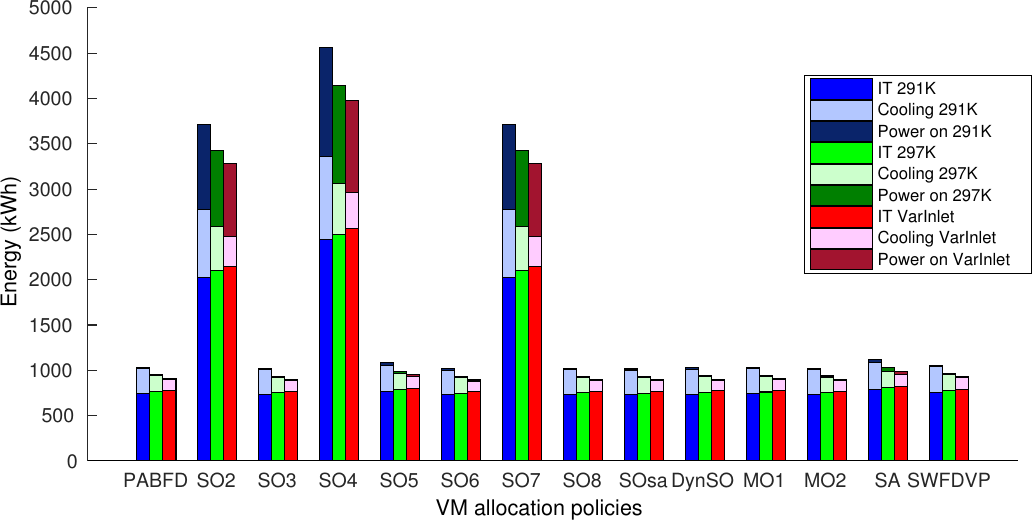}
\setlength{\belowcaptionskip}{-20pt}
\caption{Contributions to data center energy per VM allocation strategy for Workload 2}
\label{fig:figureStackedWL2}
\end{figure}

\begin{figure} [h!]
\centering
\includegraphics[width=0.6\columnwidth]{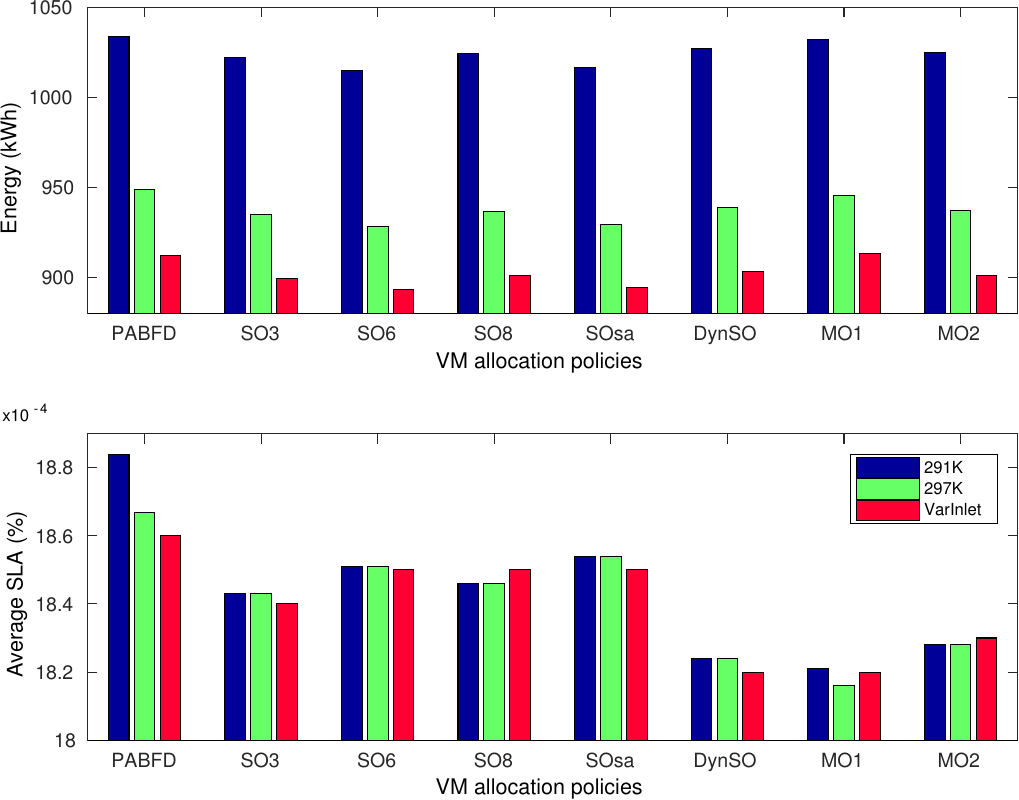}
\setlength{\belowcaptionskip}{-20pt}
\caption{Data center energy and SLA per VM allocation strategy for Workload 2}
\label{fig:nonstackedWL2}
\end{figure}

Figure~\ref{fig:nonstackedWL2} shows the energy and average SLA
percentage comparison for those strategies that outperform the
baseline PABFD, which is the baseline that performs better in this scenario, in terms of energy, where SLA is maintained. As in the Workload 1
scenarios, $SO_{SA}$, $SO_6$, $SO_3$, $MO_2$ and \textit{DynSO} allocation
policies offer the best results, in terms of energy savings, when
compared to our baselines SA, PABFD and SWFDVP. These policies, when
combined with \textit{VarInlet} strategy, provide average savings of 12.48\%
and 19.99\% with respect to SA at 297~K and 291~K respectively, and
maximum savings of up to 12.97\% and 20.43\% respectively for
$SO_{6}$.  These policies, when compared with PABFD provide average
savings of 5.34\% and 13.09\% at 297~K and 291~K respectively, and
maximum savings of up to 5.88\% and 13.59\% respectively for $SO_{6}$.
Comparing the results with the SWFDVP baseline, average savings of 7.20\% and 15.25~\% and maximum savings of 7.72~\% and 15.36\% (for $SO_{6}$) are found.

Finally, we optimize Workload 3, which presents the lowest
instantaneous variability.  For this optimization scenarios, we obtain
the energy consumption, average SLA and migration results provided in
Table~\ref{tab:WL3}.

\begin{table} [phtb]
\begin{center}
\caption{Energy, SLA and Migration metrics per inlet temperature and allocation policy for workload 3.}
\label{tab:WL3}
\begin{tabular}{lrrrrrrrrr}
\hline
\multirow{2}{*}{Policy} & \multicolumn{3}{c}{Energy (kWh)} & \multicolumn{3}{c}{Average SLA ($\cdot 10^{-4}$ \%)} & \multicolumn{3}{c}{Migrations ($\cdot 10^{3}$)} \\
& 291K & 297K & VarInlet & 291K & 297K & VarInlet & 291K & 297K & VarInlet \\
\hline
PABFD & 855.52	& 786.93	& 759.02	& 15.68	& 15.77	& 15.74	& 64.4	& 63.2	& 67.5  \\   
$SO_2$ & 2715.34	& 2505.51	& 2381.12	& 10.44	& 10.49	& 10.46	& 467.9	& 477.7	& 466.9 \\
$SO_3$ & 852.66		& 781.12	& 752.55	& 15.35	& 15.35	& 15.35	& 70.3	& 70.3	& 70.3  \\ 
$SO_4$ & 3725.87	& 3378.07	& 3232.15	& 10.37	& 10.37	& 10.38	& 718.1	& 718.1	& 721.4 \\  
$SO_5$ & 884.77		& 810.29	& 780.37	& 15.47	& 15.47	& 15.47	& 73.3	& 73.3	& 73.3  \\ 
$SO_6$ & 858.16		& 785.95	& 757.06	& 15.39	& 15.39	& 15.39	& 74.7	& 74.7	& 74.7  \\ 
$SO_7$ & 2715.34	& 2505.51	& 2381.12	& 10.44	& 10.49	& 10.46	& 467.9	& 477.7	& 466.9 \\  
$SO_8$ & 855.74		& 783.85	& 755.10	& 15.32	& 15.32	& 15.32	& 72.5	& 72.5	& 72.5  \\ 
$SO_{SA}$ & 856.41	& 84.40		& 755.58	& 15.02	& 15.02	& 15.02	& 75.7	& 75.7	& 75.7  \\ 
$DynSO$ & 853.27	& 781.65	& 753.01	& 15.21	& 15.21	& 15.21	& 72.3	& 72.3	& 72.3  \\ 
$MO_1$ & 864.66		& 787.28	& 757.73	& 15.20	& 15.26	& 15.41	& 73.1	& 71.0	& 70.1  \\ 
$MO_2$ & 859.39		& 787.07	& 758.11	& 15.21	& 15.21	& 15.21	& 73.8	& 73.8	& 73.8  \\ 
$SA$   & 946.98 	& 856.72	& 827.95	& 14.87 & 14.43	& 14.67	& 93.9	& 95.4	& 96.9  \\
SWFDVP & 888.44	& 809.07	& 808.57	& 14.67 & 14.90 & 14.93 & 70.9  & 74.5  & 71.94 \\
\hline
\end{tabular}
\end{center}
\end{table}

\begin{figure} [h!]
\centering
\includegraphics[width=0.8\columnwidth]{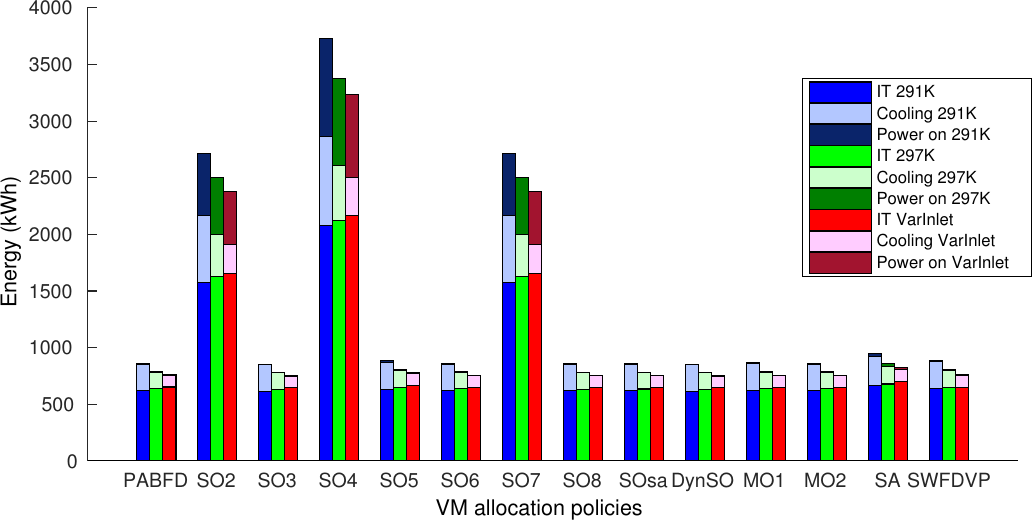}
\setlength{\belowcaptionskip}{-20pt}
\caption{Contributions to data center energy per VM allocation strategy for Workload 3}
\label{fig:stackedWL3}
\end{figure}

In Figure~\ref{fig:stackedWL3}, the same trend towards power and
temperature is presented as in Workload 1 and Workload 2 scenarios.
Increasing cooling set points provides savings that outperform IT
power increments, thus resulting in more efficient scenarios for all
the proposed allocation strategies.  For these scenarios, only our
\textit{VarInlet} cooling strategy provides additional energy savings of
3.94\% and 11.99\% in average when compared with fixed cooling at
297~K and 291~K respectively for all the allocation policies.

\begin{figure} [h!]
\centering
\includegraphics[width=0.6\columnwidth]{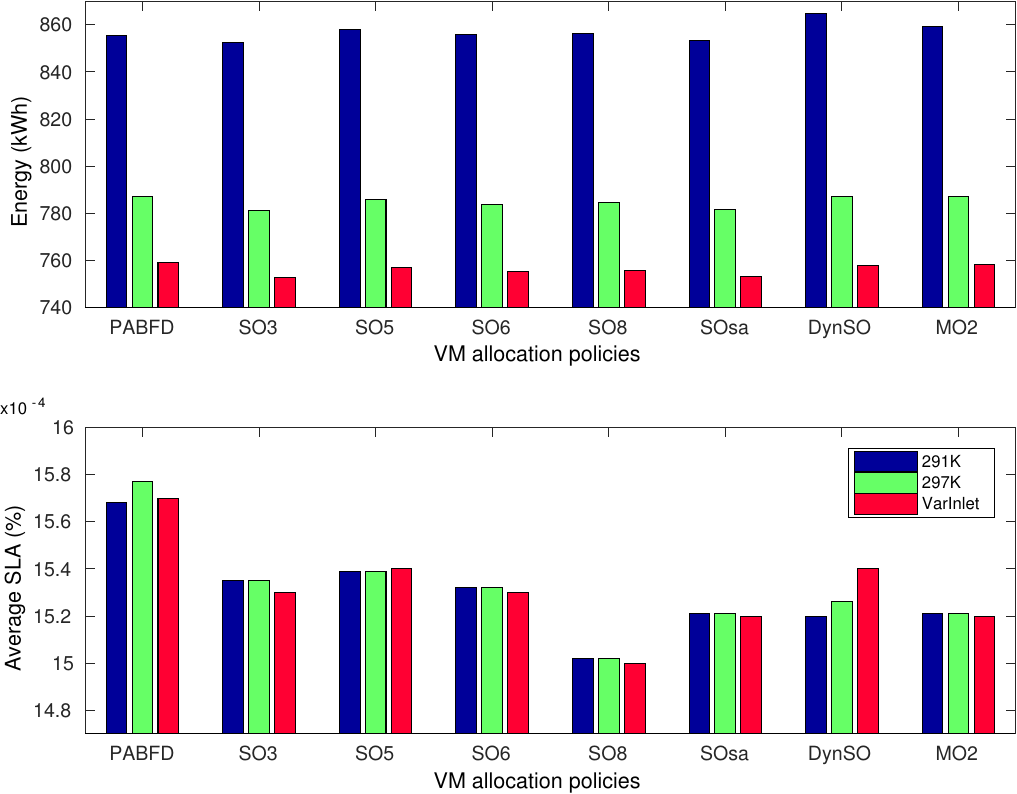}
\setlength{\belowcaptionskip}{-20pt}
\caption{Data center energy and SLA per VM allocation strategy for Workload 3}
\label{fig:nonstackedWL3}
\end{figure}

Figure~\ref{fig:nonstackedWL3} shows the energy and average SLA
percentage comparison for those strategies that outperform the
baseline PABFD, which is the baseline that performs better in this scenario, in terms of energy, where SLA is maintained. For this workload,
$SO_{SA}$, $SO_6$, $SO_3$, $MO_1$ and \textit{DynSO} allocation policies
offer the best results, in terms of energy savings, when compared to
our three baselines. These policies, when combined with \textit{VarInlet} strategy,
provide average savings of 11.78\% and 20.19\% with respect to SA at
297~K and 291~K respectively, and maximum savings of up to 12.16\% and
20.53\% respectively for $SO_{3}$.  These allocation policies, when
compared with $SO_1$ also provide average savings of 4.02\% and
11.72\% at 297~K and 291~K respectively, and maximum savings of up to
4.37\% and 12.04\% respectively for $SO_{3}$.
Finally, comparing the results with the SWFDVP baseline, average savings of 7.22\% and 15.51~\% and maximum savings of 7.53~\% and 15.79\% (for $SO_{3}$) are found.

Tables~\ref{tab:summarySavings2}, \ref{tab:summarySavings} and ~\ref{tab:summarySavings3} provide
a summary of the energy savings obtained for the different
optimization scenarios for \textit{VarInlet} cooling strategy, compared with
the local baseline PABFD, the global SA baseline  and the SWFDVP state-of-the-art baseline policies
respectively.  The results show that higher energy savings are
provided for increasing instantaneous workload variability for both
fixed cooling inlet strategies at 291~K and 297~K.
\begin{table} [htb]
\begin{center}
\caption{Energy savings for \textit{VarInlet} per VM allocation policy compared with PABFD.}
\label{tab:summarySavings2}
\begin{tabular}{lllllll}
\hline
\multirow{3}{*}{Policy} & \multicolumn{6}{c}{Energy Savings for \textit{VarInlet} vs. PABFD (\%)} \\
& \multicolumn{2}{c}{Workload 1} & \multicolumn{2}{c}{Workload 2} & \multicolumn{2}{c}{Workload 3} \\
& 297K & 291K & 297K & 291K & 297K & 291K \\
\hline
$SO_{3}$ & 6.53         & 13.60 & 5.24  & 13.00 & 4.37  & 12.04 \\
$SO_{6}$ & 6.64         & 13.71 & 5.88  & 13.59 & 3.80  & 11.51 \\
$SO_{SA}$ & 7.07        & 14.10 & 5.75  & 13.47 & 3.98  & 11.68 \\
$DynSO$ & 6.30          & 13.39 & 4.83  & 12.62 & 4.31  & 11.98 \\
$MO_{2}$ & 6.50         & 13.58 & 5.04  & 12.81 & 3.66  & 11.39 \\
\hline
\end{tabular}
\end{center}
\end{table}

\begin{table} [htb]
\begin{center}
\caption{Energy savings for \textit{VarInlet} per VM allocation policy compared with SA.}
\label{tab:summarySavings}
\begin{tabular}{cllllll}
\hline
\multirow{3}{*}{Policy} & \multicolumn{6}{c}{Energy Savings for \textit{VarInlet} vs. SA (\%)} \\
& \multicolumn{2}{c}{Workload 1} & \multicolumn{2}{c}{Workload 2} & \multicolumn{2}{c}{Workload 3} \\
& 297K & 291K & 297K & 291K & 297K & 291K \\
\hline
$SO_{3}$ & 13.59 & 21.28 & 12.38 & 19.89 & 12.16 & 20.53 \\
$SO_{6}$ & 13.70 & 21.38 & 12.97 & 20.43 & 11.63 & 20.06 \\
$SO_{SA}$ & 14.09 & 21.74 & 12.85 & 20.33 & 11.81 & 20.21 \\
$DynSO$ & 13.39	 & 21.10 & 12.00 & 19.54 & 12.11 & 20.48 \\
$MO_{2}$ & 13.57 & 21.27 & 12.19 & 19.72 & 11.51 & 19.94 \\
\hline
\end{tabular}
\end{center}
\end{table}

\begin{table} [htb]
\begin{center}
\caption{Energy savings for \textit{VarInlet} per VM allocation policy compared with SWFDVP.}
\label{tab:summarySavings3}
\begin{tabular}{lllllll}
\hline
\multirow{3}{*}{Policy} & \multicolumn{6}{c}{Energy Savings for \textit{VarInlet} vs. SWFDVP (\%)} \\
& \multicolumn{2}{c}{Workload 1} & \multicolumn{2}{c}{Workload 2} & \multicolumn{2}{c}{Workload 3} \\
& 297K & 291K & 297K & 291K & 297K & 291K \\
\hline
$SO_{3}$ & 6.82  & 14.72 & 7.09 & 14.78 & 7.53 & 15.79  \\
$SO_{6}$ & 6.94  & 14.82 & 7.72 & 15.36 & 7.02 & 15.33 \\
$SO_{SA}$ & 7.36 & 15.21 & 7.59 & 15.25 & 7.18 & 15.47 \\
$DynSO$ & 6.60   & 14.51 & 6.69 & 14.41 & 7.52 & 15.78 \\
$MO_{2}$ & 6.80  & 14.70 & 6.89 & 14.60 & 6.85 & 15.17  \\
\hline
\end{tabular}
\end{center}
\end{table}

The proposed VM allocation strategies $SO_{3}$, $SO_{6}$, $SO_{SA}$, \textit{DynSO},
$MO_{2}$ cannot be compared between them in terms of power savings, as their
relative savings fall behind the error of our power model.  In terms of SLA,
the outcomes show that our algorithms maintain the SLA obtained for the
baseline policy. The SLA violations are increased by $SO_6$ and $SO_{SA}$ for
the most variable scenario (Workload 1), and only when compared with the 291~K
fixed cooling policy. However, this increment is only of about
$0.14\cdot10^{-4}$.  If the SLA is critical for the data center management,
$MO_2$ provides SLA reductions that are consistent within the 9 scenarios, also
offering competitive energy savings.
Globally, our $SO_{SA}$ is the strategy that performs better if selected for
all the different scenarios with significantly different workload profiles.
This approach presents the best savings for Workload 1, which is the more
variable one, and a very high savings value for less variable workloads 2 and
3 (for which both baselines are more competitive due to the lower variability).  
For all the scenarios, the $SO_{SA}$ approach outperforms the PABFD, SA and SWFDVP baselines
as can be seen in
Table~\ref{tableSOSAbest}.
Our local SO based on SA optimization,$SO_{SA}$, leverages the
information from a global strategy combined with the information of the overall
data center infrastructure provided by our optimization framework.

\begin{table} [htb]
\begin{center}
\caption{Energy savings for \textit{VarInlet} in average and for $SO_{SA}$ strategy.}
\label{tableSOSAbest}
\begin{tabular}{llllllll}
\hline
\multirow{2}{*}{Baseline} & & \multicolumn{2}{c}{Workload 1} & \multicolumn{2}{c}{Workload 2} & \multicolumn{2}{c}{Workload 3} \\
& & 297K & 291K & 297K & 291K & 297K & 291K \\
\hline
\multirow{2}{*}{PABFD} & average & 	 6.61\% & 13.67\% &  5.34\% & 13.09\% &  4.02\% & 11.72\% \\
 & $SO_{SA}$ & 	 7.07\% & 14.10\% &  5.75\% & 13.47\% &  3.98\% & 11.68\% \\
\multirow{2}{*}{SA} & average & 	13.67\% & 21.35\% & 12.48\% & 19.99\% & 11.78\% & 20.19\% \\
 & $SO_{SA}$ & 	14.09\% & 21.74\% & 12.85\% & 20.33\% & 12.16\% & 20.50\% \\
\multirow{2}{*}{SWFDVP} & average & 	 6.91\% & 14.79\% &  7.20\% & 14.88\% &  7.22\% & 15.51\% \\
 & $SO_{SA}$ & 	 7.36\% & 15.21\% &  7.59\% & 15.25\% &  7.18\% & 15.47\% \\
\hline

\end{tabular}
\end{center}
\end{table}

Each workload (1 to 3) is composed of 7 days of real traces obtained each 300
seconds (2016 traces, each of them defining an optimization slot). In terms of
computational time the BFD-based approaches show a better performance.  The
simulation time for the SO policies ($SO_{1}$-$SO_{8}$, \textit{DynSO} and $SO_{SA}$) ranges from
56 to 84 minutes. $MO_1$ and $MO_2$ present a simulation time of around
105 minutes. Thus, each minimization (one per optimization slot) for BFD-based
approaches takes between 1.66 and 3.12 seconds.  On the other hand, the
complete simulation of a 7 days-workload takes between 3360 and 5040 minutes to
finish for the SA scenario. So, each SA minimization takes between 100 and 150
seconds approximately, and as our SA is configured with 100000 iterations, each
iteration takes only 1 millisecond. These results are compliant with the space
of solutions explored in each case, \textit{numVMs*numHosts=1127*1200} for BFD-based and
$numHosts^{numVMs}=1200^{1127}$ for the SA-based approaches.
Thus, these results show that the BFD-based algorithms are a
lighter implementation in terms of execution time and complexity, but still
powerful enough to tackle the joint IT and cooling energy optimization problem like the one
described in this research.

Finally, the PUE (Power Usage Effectiveness) is a metric that measures the
efficiency of a data center in terms of energy usage. Specifically, it provides
information of the amount of energy that is used by the computing equipment in
contrast to cooling and other overheads.  The PUE value works well with cooling
optimizations, as reductions on cooling power result in a higher percentage of
the power budget consumed only by IT. On the other hand, for IT optimizations
that do not impact on cooling consumption, the PUE does not show the efficiency
gained by the IT energy reduction, but reports a negative impact, as it results
on a higher percentage of the total power used for cooling purposes.  Our research
reduces both IT and cooling contributions to final power consumption
simultaneously. For fixed cooling baseline policies with set point temperatures of 291~K
and 297~K, our infrastructure provides PUE values of 1.37 and 1.23
respectively.  Our \textit{VarInlet} approach, together with our proposed dynamic
consolidation $SO_{SA}$, optimizes the PUE in up to 16.05\% and
6.5\% respectively, providing a PUE value of 1.16, for a dynamic utilization of
the IT resources around 80\%, that outperforms the state-of-the-art value that
is around 1.2.

\section{Conclusions}
\label{sec:conclusionsTemp}
The new optimization framework proposed in this work focuses on considering
the energy globally from the data center perspective. In this way, the elements are aware
of the evolution of the global energy demand and the thermal behavior
of the room.  Our decisions are taken based on information from the
available subsystems to perform energy optimizations from the
technology to the data center level.
Metaheuristic algorithms like Simulated Annealing, when used for VM
consolidation in data centers, are able to achieve very good results
in terms of energy.  However, the time that they need to perform the
optimizations makes them unsuitable for being used during runtime for
this purpose.  On the other hand, various local BFD-based policies
provide good solutions to the energy problem. They constrain the set
of active servers, thus reducing the static energy consumption, but
this local strategies do not consider the final status of the data
center after each optimization, so the number of VM migrations may be
increased.
Leveraging the knowledge gathered from both metaheuristic and BFD
algorithms helps us to infer models that describe global energy
patterns into local strategies, which are faster and lighter to be
used for optimizing the energy consumption during runtime under
variable conditions. By using this technique we provide the $SO_{SA}$
VM allocation policy that, together with our proposed cooling strategy
$VarInlet$, allows us to improve the energy efficiency in scenarios
with high workload variability.  
Our $SO_{SA}$ technique achieved
energy savings around 7\% and 15\%, 
using two different traditional cooling strategies with
fixed set points at 297~K and 291~K respectively,
when compared with both local
baselines PABFD and SWFDVP. 
Also, compared with
a global SA-based baseline, our local VM allocation policy $SO_{SA}$ provided
energy savings of up to 14.09\% and 21.74\%. For all the scenarios
proposed in this research, our optimization algorithms maintain the
QoS when compared with a local and a global baseline.

\section*{Acknowledgment}
\label{sec:acknowledgment}
This project has been partially supported by the EU (FEDER), the Spanish
Ministry of Economy and Competitiveness, under contracts TIN-2015-65277-R,
TEC-2012-33892, AYA2015-65973-C3-3-R and RTC-2014-2717-3.

\bibliographystyle{apalike}
\bibliography{SPE}

\end{document}